\documentclass[journal]{IEEEtran}
\usepackage{cite}
\usepackage{amsmath,amssymb,amsfonts}
\usepackage{algorithmic}
\usepackage{graphicx}
\usepackage{textcomp}
\usepackage{harpoon}
\usepackage{amsmath}
\usepackage{mathrsfs}
\usepackage{amssymb}
\usepackage{epstopdf}
\usepackage{subfig}
\usepackage{lmodern}
\usepackage{bm}
\usepackage{sansmath}
\usepackage{booktabs}
\usepackage{colortbl}
\usepackage{esint}
\usepackage{stfloats}
\usepackage{lipsum}
\usepackage{multicol}
\usepackage{multirow}
\usepackage{tabularx}
\usepackage{color}

\newenvironment{proof}{{\textsl {Proof:}}}{\hfill$\blacksquare$\par}

\ifCLASSINFOpdf

\else

\fi

\usepackage[absolute,showboxes]{textpos}

\TPMargin{3pt}
\newcommand{\copyrightstatement}{
	\begin{textblock}{0.84}(0.08,0.95) 
		\noindent
		\footnotesize
		\copyright 2023 IEEE. Personal use of this material is permitted. Permission from IEEE must be obtained for all other uses, in any current or future media, including reprinting/republishing this material for advertising or promotional purposes, creating new collective works, for resale or redistribution to servers or lists, or reuse of any copyrighted component of this work in other works. DOI: 10.1109/JIOT.2023.3313870.
	\end{textblock}
}

\begin{document}
\copyrightstatement
\title{Joint Information and Jamming Beamforming for Securing IoT Networks With Rate-Splitting}

\author{Bin~Qiu,~\IEEEmembership{Member,~IEEE,}
        Wenchi~Cheng,~\IEEEmembership{Senior Member,~IEEE,}
        and~Wei~Zhang,~\IEEEmembership{Fellow,~IEEE}

\thanks{This work was supported in part by the National Key R\&D Program of China under Grant 2021YFC3002102, in part by the Key R\&D Plan of Shaanxi Province under Grant 2022ZDLGY05-09, in part by the Key Area R\&D Program of Guangdong Province under Grant 2020B0101110003, in part by the Fundamental Research Funds for the Central Universities under Grant XJS220105, in part by the Project funded by China Postdoctoral Science Foundation under Grant 2022M712491, and in part by the Natural Science Basic Research Program of Shaanxi under Grant 2023-JC-QN-0715. (Corresponding author: Wenchi Cheng.)}

\thanks{Bin Qiu and Wenchi Cheng are with the State Key Laboratory of Integrated Services Networks, Xidian University,
	Xian 710071, China (e-mail: qiubin@xidian.edu.cn; wccheng@xidian.edu.cn).}

\thanks{Wei Zhang is with the School of Electrical Engineering and Telecommunications, University of New South Wales, Sydney, NSW 2052, Australia (e-mail: w.zhang@unsw.edu.au).}
}

\maketitle

\begin{abstract}
The goal of this paper is to address the physical layer (PHY) security problem for multi-user multi-input single-output (MU-MISO) Internet of Things (IoT) systems in the presence of passive eavesdroppers (Eves). To this end, we propose an artificial noise (AN)-aided rate-splitting (RS)-based secure beamforming scheme. Our design considers the dual use of common messages and places the research emphasis on hiding the private messages for secure communication. In particular, leveraging AN-aided RS-based beamforming, we aim to maximize the focused secrecy sum-rate (F-SSR) by jointly designing transmit information and AN beamforming while satisfying the desired received constraints for the private messages at IoT devices (IoDs), and per-antenna transmit power constraint at base station. Then, we proposed a two-stage algorithm to iteratively find the optimal solution. By transforming non-convex terms into linear terms, we first reformulate the original problem as a convex program. Next, we recast the optimization problem to an unconstrained problem to obtain the global optimal solutions. Utilizing the duality framework, we further develop an efficient algorithm based on a barrier interior point method to solve the reformulated problem. Simulation results validate the superior performance of our proposed schemes.
\end{abstract}

\IEEEpeerreviewmaketitle

\section{Introduction}
\IEEEPARstart{T}{he} sixth-generation (6G) wireless communication networks are envisioned to revolutionize customer services and applications via the Internet of Things (IoT) to a future of highly intelligent and autonomous system \cite{6G_Internet}. With the dramatic increase of IoT devices (IoDs) in a variety of emerging application scenarios, such as smart city, data analysis, intelligent transportation, and security surveillance, vast amounts of private and resource information are interacted through IoT networks \cite{Scalable_Fu,Disaster,Latency_Cheng}. Similarly to other wireless networks, IoT networks are particularly faced with security threats due to the broadcasting kind of wireless medium. Unfortunately, many low-end IoT commercial products do not usually support strong security mechanisms, and can hence be target of a number of security attacks. Traditionally, secure communication relies on the cryptographic encryption. However, it may fail to provide satisfactory secure transmission with the development of computing power and intelligent detection technologies in future networks. Physical layer (PHY) security, as an alternative method, which does not depend on the computing power and key management, has attracted extensive attention recently \cite{Survey_Wang}. For PHY security, the key is to exploit the randomness of wireless channels to achieve encrypting data transmission \cite{secure_JCIN}. Specifically, PHY security technique allows the enhancement of signal gain on the desired users' directions while reducing power leakage and/or debilitating signal phase along the eavesdropper (Eve) by judiciously designing the beamforming of the multiple transmit antennas \cite{Symbol_Kalantari}. 

It is of interest to exploit flexibility at PHY that provides secure wireless communication. The wiretap channel is a fundamental primitive to model eavesdropping at the PHY \cite{wiretap_Wyner}. After that,  Csis\'{z}ar and K\"{o}rner studied the secure communication over broadcast channel \cite{Broadcast_Csiszar}, and established the secrecy rate, which is an epoch-making metric of measuring the security performance. Inspired by these works, nowadays various techniques have been researched to achieve PHY security. In \cite{Switched_Alotaibi}, the authors proposed switched phased-array transmission schemes to scramble the constellation in both amplitude and phase along undesired directions. The authors of \cite{Polygon_Zhang} considered a phased-array transmission structure via polygon construction PSK modulation to achieve secure transmission. Also, it is highly promising to use PHY security in IoT networks. By maximizing the secrecy sum rate, the authors in \cite{Cognitive_Radio_IoT} jointly designed the beamforming matrix and vectors for a two-way cognitive radio IoT networks.

Additionally, jamming noise was frequently used to improve the PHY security performance by degrading the channel condition of the eavesdropping link, which is called as artificial noise (AN). Aware that intentional jamming is able to reduce eavesdroppers’ capabilities, the authors of \cite{Artificial_Goel} were first investigated an AN-aided beamforming scheme that consumes a certain transmit power to generate AN so as to hide information transmission. The authors of \cite{NOMA-Based_IoT} presented an AN cooperative transmission scheme for  non-orthogonal multiple access (NOMA)-based IoT networks. Additionally, the authors of \cite{Precise_Shu} employed multiple tools—random subcarrier, AN, beamforming, and OFDM to achieve an ultimate aim of PHY security communication. Nevertheless, the conventional transmit beamforming and AN-aided design fail to provide PHY security when Eves fall within the main-lobe beam due to the limitation of the only angle-dependent characteristic of the beampattern \cite{Hybrid}.

Rate-splitting multiple access (RSMA) has emerged as a novel, general, and powerful framework for the design and optimization of non-orthogonal transmission, multiple access, and interference management strategies for future wireless networks \cite{Survey_Mao}. Under the RSMA umbrella, various rate-splitting (RS) architectures have been developed. By partially decoding and partially treating interference as noise, RS can result in spectral efficiency \cite{Spectral_Zhou}, energy efficiency \cite{Energy_Mao}, robustness \cite{Robust_Joudeh}, and security enhancement  \cite{Secure_Splitting} over wireless networks. For overloaded cellular IoT networks, the authors in \cite{Overloaded_Cellular}  analyzed  the degrees of freedom and  rate. Two new multiple access techniques based on multi-antenna RS, time partitioning–RSMA and power partitioning–RSMA, were proposed to achieve the optimal degrees of freedom (DoF). RS provides new ideas for PHY security by multiplex common and private beams. By dual using energy of common message as transmit information and interference, the authors of \cite{Robust_Fu} developed a robust secure beamforming design method to maximize the worst-case secrecy rate in multi-input single-output (MISO) systems. In \cite{UAV_Bastami}, the max-min secrecy fairness of cellular networks was investigated, in which cooperative RS aided down-link transmissions are employed to safeguard the downlink of a two-user system against an external multi-antenna Eves. Besides, the authors in \cite{Rate_Cognitive} proposed an application of RS by joint communications and jamming under a multi-carrier waveform for a multi-antenna cognitive radio system. It is worth noting that most of the prior research on PHY security directly maximized the achievable secrecy rate. However, the achievable secrecy rate requires the perfect or estimate of location information of Eves. It may not be possible to acquire any information of the passive devices in practice.

Motivated by the aforementioned aspects and secure performance enhancement requirements,  a new flexibility of beamforming transmission technique is urgent to futher enhance the PHY security for IoT networks. In this paper, we conceive an AN-aided RS-based secure communication scheme for the multi-user multiple input single output (MU-MISO) IoT networks in the presence of multiple passive Eves. The main contributions of this paper are listed as below:

\hangindent 2.3em
1) We pioneer the study of the application of AN-aided RS-based beamforming framework in IoT secure transmission systems with multiple passive Eves. We maximize the focused secrecy sum-rate (F-SSR) by jointly designing the transmit information and AN beamforming subject to the received signal-to-noise ratio (SNR) constraints of the private streams and the per-antenna transmit power constraints. In this way, the PHY security is enhanced by the dual use of the common message for RS, which is actually serving both as a desired message and interference for IoDs and Eves, respectively. In particular, private messages are implicitly embedded in the common messages. 

\hangindent 2.3em
2) To fill the research gap on the practical implementation, our design scheme is under the case of statistical sense of Eves' channel \cite{Joint_Nguyen} due to the passive nature of Eves. Additionally, since each antenna is equipped with its own power amplifier and is limited individually by the linearity of the amplifier in practice, we employ a more realistic per-antenna power constraint.

\hangindent 2.3em
3) To handle the F-SSR maximization problem, we first introduce an auxiliary variable to confine the allowable signal-to-interference-plus-noise ratio (SINR) of private parts for Eves. Then, we present a two-stage algorithm to make the problem feasible. In particular, the auxiliary variable is fixed in the first stage. The non-convex terms of the objective are transformed into linear terms, and then the semidefinite programming (SDP) relaxation approach can be applied for the suitable reformulation; next, in the second stage, the global solution of the problem is obtained via the Broyden-Fletcher-Goldfarb-Shanno (BFGS) method \cite{BFGS_Nezhad}, which is a Lagrange dual program related to the quasi-Newton optimization. Utilizing the special features, the considered problem is reduced to a mini-max program via duality. It facilitates a barrier interior point method to obtain the optimal solution.

The rest of this paper is organized as follows. Section II describes the system model of the AN-aided RS-based beamforming transmission and formulates the optimization problem. Some insights into the F-SSR maximization problem is provided; then, a two-stage algorithm is proposed to solve the problem  in Section III. An efficient mini-max program to solve the problem is extended in Section IV. Simulation results are shown in Section V. Finally, the conclusion is drawn in Section VI.

\section{System Model and Problem Formulation}
In this section, we specify the system model of AN-aided RS-based beamforming for secure communication followed by the formulated F-SSR maximization problem.
\subsection{Transmit Signal Model}
\begin{figure}
	\centering
	\includegraphics[width=1\columnwidth]{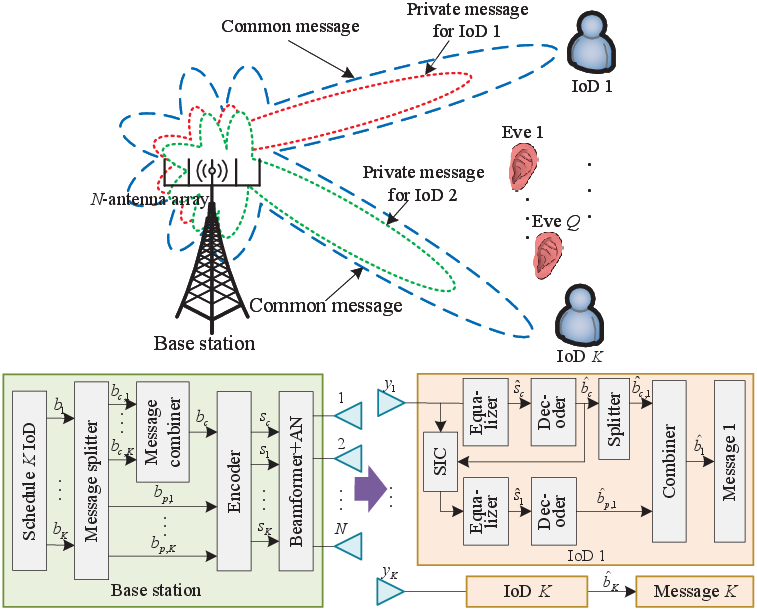}
	\caption{System model of AN-aided RS-based secure beamforming transmission.}
	\label{fig1}
\end{figure}

We consider a MU-MISO IoT transmission system that consists of an array central controller/base station, $K$ single-antenna IoDs, and $Q$ single-antenna passive Eves, $\footnote{{Given the passive nature of Eves, the information of the detection channel is hard to be precisely acquired by base station, including the number. For analytical tractability, we assume a particular number for $Q$ such that the based station can handle at most $Q$ Eves.}}$ as illustrated in Fig. \ref{fig1}. The base station equipped with $N$ isotropic antennas provides wireless service to IoDs whereas the Eves try to eavesdrop. Following the RS principle \cite{Gaussian_Rimoldi}, we employ multi-antenna RS at base station and successive interference cancellation (SIC) at the IoDs. More specially, the confidential messages $b_k$ intended to IoD $k$, $k \in \mathcal{K}$, ${\cal K}= \{ 1,2,...K\}$ are split into common parts and private parts. The common parts are combined into common messages $b_{c,k}$, $k \in \mathcal{K}$, which are packed into the common stream $s_c$ shared to all IoDs. The private messages $b_{p,k}$ are independently encoded into the private streams $s_k$ sent to IoD $k$, $k \in \mathcal{K}$. As a result, we group the transmit symbols as a vector, i.e., ${\bf{s}} = {\left[ {{s_c},{s_1}, \cdots,{s_K}} \right]^T} \in {\mathbb{C}^{(K + 1) \times 1}}$ with $\mathbb{E}\left\{ {{\bf{s}}{{\bf{s}}^H}} \right\} = {{\bf{I}}_{K + 1}}$, where $\mathbb{E}\{\cdot\}$ indicates the expectation. The private streams are mapped to transmit array in a multicast fashion, while the common stream beamforming is designed in a broadcast manner. As a consequence, the AN-aided RS-based beamforming transmit baseband signal, denoted by ${\bf{x}}$, is described as \cite{Hybrid, Robust_Fu}
\begin{eqnarray}
 {\bf{x}} = {{\bf{w}}_c}{s_c} + \sum\limits_{ k \in \mathcal{K}} {{{\bf{w}}_k}{s_k}}  + {{\bf{n}}_a},
\label{eq1}\end{eqnarray}
where ${{\bf{w}}_c}\in \mathbb{C}^{N \times 1}$ and ${{\bf{w}}_k}\in \mathbb{C}^{N \times 1}$, $k \in \mathcal{K}$, indicate the beamforming to control the common stream ${s_c}$ and the private streams ${s_k}$, respectively, ${\bf{n}}_{a}\in \mathbb{C}^{N \times 1}$ is the AN, whose elements satisfy ${{\mathbf{n}}_{a}}={{\bf{P}}}{{\bf{z}}}$, ${{\bf{P}}}\in \mathbb{C}^{N \times (N-K)}$ is the AN projection matrix for imposing a disturbance to Eves, ${\mathbf{z}}\in \mathbb{C}^{(N-K) \times 1}$ is an AN vector, which consists of complex Gaussian variables with zero-mean and unit-variance, satisfying ${\bf{z}}\sim{\mathcal{CN}}({{{0}}},{{\bf{I}}_{N-K}})$. 

Without loss of generality, we employ an uniform linear array (ULA) consisted of isotropic antennas at base station, and the features can be easily applied to multi-dimensional periodic arrays. The first element of the ULA is viewed as the origin of the coordinate system and the phase reference. Let $d$ denote the ULA' s adjacent elements spacing, satisfying $d = c/(2f_c)$ to avoid creating grating lobes, where $f_c$ and $c$ indicate the carrier frequency and the speed of light, respectively. All involved channels are modeled as far-field line-of-sight (LoS) transmission, and the LoS assumption also captures the essence to facilitate the ongoing and fruitful high-frequency secure communications. $\footnote{Following the spirit of \cite{Polygon_Zhang}, the channel model can be extended to the Saleh-Valenzuela geometric model with multi-path mmWave transmission.}$ Therefore, the instantaneous received signal for IoDs at time $t$, denoted by $	y\left( {r,\theta; t} \right)$, represents as
\begin{eqnarray}
\begin{aligned}[b]
	y\left( {r,\theta ;t} \right) & = \sum\limits_{ n \in \mathcal{N}} {\rho \left( {{r}} \right)}{e^{j2\pi {f_c}\left[ {t - \frac{{r - \left( {n - 1} \right)d\sin \theta }}{c}} \right]}} {\left[ {\bf{x}} \right]_n} + {n_c}\\&
		=  {\rho \left( {{r}} \right)}{e^{j2\pi {f_c}\left( {t - \frac{r}{c}} \right)}}\sum\limits_{ n \in \mathcal{N}} {\Phi_{n} \left( \theta \right)} {\left[ {\bf{x}} \right]_n} + {n_c},
\end{aligned}
\label{eq2}\end{eqnarray}
where ${[ \cdot ]_{i}}$ denotes the $i$th element of the vector, $\Phi_{n} \left( \theta \right) = {e^{\frac{{j2\pi {f_c}\left( {n - 1} \right)d\sin \theta }}{c}}}$, $\forall n \in \mathcal{N}$, ${\cal N} = \{ 1,2,...N\}$, $\rho \left( {{r}} \right)$ denotes the signal attenuation factor, $r$ denotes the base station-IoD distance, $\theta$ is the direction of IoD, and $n_c$ is the background thermal noise. Due to the far-field transmission, satisfying $r \gg d$, the attenuation difference among the transmit antennas is negligible.

Let us use ${\bf{h}}\left(r,\theta \right)$ to denote the channel vector of the links from the transmit array to IoD as
\begin{eqnarray}
{\bf{h}}\left(r,\theta \right) = \rho (r){[{{{\Phi _1}\left( \theta  \right)}},{{{\Phi _2}\left( \theta  \right)}},...,{{{\Phi _N}\left( \theta  \right)}}]^H}\in {\mathbb{C}^{N \times 1}}.
\label{eq3}\end{eqnarray}

Denote by $\left( {{r_{u,k}},{\theta _{u,k}}} \right)$ and $\left( {{r_{e,q}},{\theta _{e,q}}} \right)$ the coordinates of the IoD $k$, $k \in \mathcal{K}$, and Eve $q$, $q \in \mathcal{Q}$, ${\cal Q}  = \{ 1,2,...Q\}$, respectively. For notational convenience, let ${\bf{h}}_{u,k}$ and ${\bf{h}}_{e,k}$ indicate the channels of the IoD $k$ and Eve $q$, i.e., ${\bf{h}}_{u,k}={\bf{h}}\left(r_{u,k},\theta_{u,k} \right)$, $\forall k \in \mathcal{K}$, and ${\bf{h}}_{e,q}={\bf{h}}\left(r_{e,q},\theta_{e,q} \right)$, $\forall q \in \mathcal{Q}$. Assume that all reception processes for IoDs are perfectly synchronized in time and frequency. Then, the received baseband signals after down-conversion at the IoD $k$ and Eve $q$, denoted by ${y_{u,k}}$ and ${y_{e,q}}$, are given by
\begin{eqnarray}
{y_{u,k}} = {\bf{h}}_{u,k}^H{\bf{x}} + {n_{u,k}}, \forall k \in \mathcal{K},
	\label{eq4}\end{eqnarray}
and
\begin{eqnarray}
{y_{e,q}} = {\bf{h}}_{e,q}^H{\bf{x}} + {n_{e,q}},  \forall q \in \mathcal{Q},
	\label{eq5}\end{eqnarray}
respectively, where ${n_{u,k}}$ and ${n_{e,q}}$ denote the additive Gaussian noise (AWGN) at the IoD $k$ and Eve $q$, satisfying ${{{{n}}_{u,k}}}\sim{\mathcal{CN}}({{{0}}},\sigma _{u,k}^2)$ and ${{{{n}}_{e,q}}}\sim{\mathcal{CN}}({{{0}}},\sigma _{e,q}^2)$, respectively.

\textsl{Remark 1:} Using a large-scale array transmitter, it has a higher spatial DoF. Hence, the beamforming technique seems to be a good candidate of security enhancement and large numbers of IoDs applications, especially with a large-scale antenna array.
\subsection{Problem Statement}
Following the processes of RS-based reception, each IoD first decodes the common streams by regarding the private streams as interference. After successfully decoding and extracting the common streams via SIC, the IoD detects its intended private streams by viewing the private streams of other IoDs as noise \cite{Robust_Fu}.  It is built upon RS, a low-complexity strategy that relies on SIC at each IoD. The corresponding capacity achieved by each IoD is directly related to its received SINR values. Given the above way, the received SINRs associated with common and private streams for IoD $k$, denoted by $\gamma _{c,k}^u$ and $\gamma _{p,k}^u$, are specified as
\begin{eqnarray}
\left\{ \begin{array}{l}
	\gamma _{c,k}^u = \frac{{{{\left| {{\bf{h}}_{u,k}^H{{\bf{w}}_c}} \right|}^2}}}{{\sum\limits_{ i \in \mathcal{K}} {{{\left| {{\bf{h}}_{u,k}^H{{\bf{w}}_i}} \right|}^2}}  + {{\left| {{\bf{h}}_{u,k}^H{{\bf{n}}_a}} \right|}^2} + \sigma _{u,k}^2}}, \forall k \in \mathcal{K},\\
	\gamma _{p,k}^u = \frac{{{{\left| {{\bf{h}}_{u,k}^H{{\bf{w}}_k}} \right|}^2}}}{{\sum\limits_{i \in {\mathcal{K}}\backslash k} {{{\left| {{\bf{h}}_{u,k}^H{{\bf{w}}_i}} \right|}^2}}  + {{\left| {{\bf{h}}_{u,k}^H{{\bf{n}}_a}} \right|}^2} + \sigma _{u,k}^2}}, \forall k \in \mathcal{K}.
\end{array} \right.
	\label{eq6}\end{eqnarray}

Under the Gaussian channel, the corresponding achievable rates of the IoD $k$ in decoding the common messages and its corresponding private messages, denoted by $R_{c,k}^u$ and $R_{p,k}^u$, can be attained by
\begin{eqnarray}
\left\{ \begin{array}{l}
	R_{c,k}^u = {\log _2}( {1 + \gamma _{c,k}^u} ), \forall k \in \mathcal{K},\\
	R_{p,k}^u = {\log _2}( {1 + \gamma _{p,k}^u} ), \forall k \in \mathcal{K}.
\end{array} \right.	
	\label{eq7}\end{eqnarray}

In contrast, the received signals at Eves are the mixed signals containing the common and all private streams. This results in mutual interference between common and all private streams. Therefore, the received SINRs of the common stream and the private stream $k$ at the Eve $q$, denoted by $\gamma _{c,q}^e$ and $\gamma _{p,q,k}^e$, can be expressed as
\begin{align}
	\left\{ \begin{array}{l}
		\gamma _{c,q}^e = \frac{{\left| {{\bf{h}}_{e,q}^H{{\bf{w}}_c}} \right|}^2}{{\sum\limits_{ k \in \mathcal{K}} {{{\left| {{\bf{h}}_{e,q}^H{{\bf{w}}_k}} \right|}^2}}  + {{\left| {{\bf{h}}_{e,q}^H{{\bf{n}}_a}} \right|}^2} + \sigma _{e,q}^2}}, \forall q \in \mathcal{Q},\\
		\gamma _{p,q,k}^e = \frac{{\left| {{\bf{h}}_{e,q}^H{{\bf{w}}_k}} \right|}^2}{{\sum\limits_{i \in {\mathcal{K}}_c \backslash k} {{{\left| {{\bf{h}}_{e,q}^H{{\bf{w}}_i}} \right|}^2}}  + {{\left| {{\bf{h}}_{e,q}^H{{\bf{n}}_a}} \right|}^2} + \sigma _{e,q}^2}},\\ \kern 35pt  \forall q \in \mathcal{Q}, \forall k \in \mathcal{K}.
	\end{array} \right.
	\label{eq8}\end{align}
where ${\mathcal{K}_c} = \mathcal{K} \cup \left\{ c \right\}$. The corresponding achievable rates of Eve $q$ in decoding the common message and the private message, denoted by $R_{c,q}^e$ and $R_{p,q,k}^e$, can be calculated by
\begin{eqnarray}
	\left\{ \begin{array}{l}
		R_{c,q}^e = {\log _2}( {1 + \gamma _{c,q}^e} ), \forall q \in \mathcal{Q},\\
		R_{p,q,k}^e = {\log _2}( {1 + \gamma _{p,q,k}^e} ), \forall q \in \mathcal{Q}, \forall k \in \mathcal{K}.
	\end{array} \right.	
	\label{eq9}\end{eqnarray}

For simplify, the thermal noise is assumed identical for all IoDs and Eves due to the similar environment and hardware architectures, i.e., $\sigma _{u}^2=\sigma _{u,k}^2, \forall k \in \mathcal{K}$, and $\sigma _{e}^2=\sigma _{e,q}^2,  \forall q \in \mathcal{Q}$. 

\textsl{Remark 2:} By designing the message split and the power allocation to the common and private streams, RS manages to partially decode the messages and views the remaining messages as interference. This capability allows RS to act as a bridge between the two extreme message management ways of fully treating messages as interference and fully decoding interference, and creates the opportunity to enhance the secure performance.

For its simplicity and effectiveness, joint design of information and AN beamforming in PHY security provisioning is desired. Furthermore, only if the decoding of both the common and private parts is correct, the confidential messages can be effectively recovered. To unleash the potential of information,  one interesting PHY security idea is to hide the private messages deep into the common messages so that the Eves' ability to wiretap the private part is degraded. Toward this end, the F-SSR maximization problem is then formulated as
\begin{align}
	\textbf{P1:}  \kern 4pt &\mathop {\max }\limits_{{{\bf{w}}_c},\left\{ {{{\bf{w}}_k}} \right\}_1^K,{{\bf{n}}_a}}  \kern 2pt \sum\limits_{ k \in \mathcal{K}} \left(R_{c,k}^u - \mathop {\max }\limits_{q \in \mathcal{Q}}\kern 2pt	R_{p,q,k}^e \right) \tag{10a} \label{eq10a}\\
	&{\rm{s.t.}} \kern 2pt {\bf{h}}_{u,k}^H{{\bf{n}}_a} = 0, \forall k \in \mathcal{K},\tag{10b} \label{eq10b}\\
	&	\kern 13pt	\sum\limits_{ i \in \mathcal{K}_c} {{{\left[ {{{\bf{w}}_i}{\bf{w}}_i^H} \right]}_{n,n}}}  + {\left[ {{{\bf{n}}_a}{\bf{n}}_a^H} \right]_{n,n}} \le {P_n}, \forall n \in \mathcal{N},\tag{10c} \label{eq10c}\\
	&	\kern 15pt 	\gamma _{p,k}^u \ge {\Gamma  _{p,k}}, \forall k \in \mathcal{K}, \tag{10d} \label{eq10d}
\end{align}
where ${[ \cdot ]_{i,j}}$ denotes the entry in the $i$th row and $j$th column of a matrix. The objective in \eqref{eq10a} is to maximize the sum achievable rate of common messages for IoDs, while minimizing the achievable rate of the private message for Eves based on the dual use of common streams as well as focused protection of the private messages.  In this way, the low-power private stream is embedded into the high-power common stream. The constraint in \eqref{eq10b} is to eliminate the AN interference with IoDs. In \eqref{eq10c}, $P_n$ denotes the $n$th transmit antenna power constraint. Due to the own power amplifier in the analog front-end for each physical implementation antenna, the transmit power need to be limited within the linearity of the power amplifier \cite{Per_Yu}. Therefore, it is more realistic to impose power constraint on a per-antenna basis. $\footnote{{In fact, our proposed per-antenna constraint can also be tuned to support the sum power constraint by modifying the constraint in \eqref{eq10c} as $\sum\limits_{i \in \mathcal{K}_c} {{\rm{Tr}} ( {{{\bf{w}}_i}{\bf{w}}_i^H} )}  + {\rm{Tr}} (  {{{\bf{n}}_a}{\bf{n}}_a^H} ) \le P_{\rm{Tol}}$, $P_{\rm{Tol}}=\sum\limits_{n \in \mathcal{N}}{P_n}$, where {\rm{Tr}}$(\cdot)$ means the trace of a matrix.}}$ In \eqref{eq10d}, ${\Gamma _{p,k}}$ is the minimum required received SINR of the private message for the IoD $k$, $\forall k \in \mathcal{K}$. The constraint is to protect the private message reception, so that the received SINR of the private streams at the IoD $k$ is more than a given threshold. 
\section{A Two-stage Algorithm to Solve F-SSR Maximization Problem}
The F-SSR maximization problem \textbf{P1} is challenging for the non-concave objective. To solve the problem, we first give some insights into \textbf{P1}, and then we propose a two-stage algorithm.
\subsection{Some Insights to the F-SSR Maximization Problem} 
Introducing an intermediate variable ${\Gamma _{e}}$ \cite{QoS_Liao}, \textbf{P1} can be equivalently solved by following problem
\begin{align}
\textbf{P2:}  \kern 4pt &\mathop {\max }\limits_{\left\{ {{{\bf{w}}_i}} \right\}_{i \in \mathcal{K}_c},{{\bf{n}}_a},{\Gamma _{e,k}}} \kern 2pt \sum\limits_{ k \in \mathcal{K}} \left[R_{c,k}^u - {{{\log }_2}\left( {1 + {\Gamma _{e}}} \right)}\right]\tag{11a} \label{eq11a}\\
	& {\rm{s.t.}} \kern 2pt\mathop {\max }\limits_{q \in \mathcal{Q}}\kern 2pt	\gamma _{p,q,k}^e \le {\Gamma _{e}}, \forall k \in \mathcal{K},\tag{11b} \label{eq11b}\\
	&
	\kern 17pt 	 \eqref{eq10b}, \eqref{eq10c}, \eqref{eq10d}, \tag{11c} \label{eq11d}
\end{align}
where ${\Gamma _{e}} > 0$ denotes maximum allowable SINR for successfully wiretapping the private streams at Eves. 

To further simplify \textbf{P2}, let us first define the channel matrix of all IoDs as
\begin{eqnarray}
	\setcounter{equation}{12}
	\begin{aligned}[b]
		{{\bf{H}}_{u, \rm{Tot}}}{{ \buildrel \Delta \over = }}\left[{\bf{h}}_{u,1},{\bf{h}}_{u,2},...,{\bf{h}}_{u,{K}}\right],
	\end{aligned}
	\label{eq12}\end{eqnarray}
where $\buildrel \Delta \over = $ denotes the definition operations. The projection matrix is placed in the null space of all the IoD channels, i.e., ${{\bf{H}}_{u, \rm{Tot}}}{{\bf{P}}}={\boldsymbol{0}}$. We perform the singular-value decomposition (SVD) operation on the IoD channel matrix, i.e., ${{\bf{H}}_{u, \rm{Tot}}^H} = {\bf{U}}[ {\begin{array}{*{20}{c}}\boldsymbol{\Sigma} &{\boldsymbol{0}}\end{array}} ]{[ {\begin{array}{*{20}{c}}{{{\bf{V}}_1}}&{{{\bf{V}}_0}}\end{array}} ]^H}$. Then, the variables can be directly modified into ${\bf{P}} \buildrel \Delta \over =  {{\bf{V}}_0}{\bf{D}}$, ${\bf{D}} \in {\mathbb{C}^{\left( {N - K} \right) \times \left( {N - K} \right)}}$. 

The information of IoD is assumed to be perfectly acquired at base station. Nevertheless, all Eves remain radio silent to hide their presence. Therefore, the information of passive Eves is not available by base station. Assume that Eve channels undergo independent and identically distributed (i.i.d.) Rayleigh fading \cite{WIPT_Ng}. A standard optimization problem \textbf{P2} is given by
\begin{align}
	\textbf{P3:} \kern 2pt & \mathop {\max }\limits_{{{\bf{W}}_i} \succeq \boldsymbol{0},{\bf{B}} \succeq \boldsymbol{0},{\Gamma _e}}  \kern 2pt {\sum\limits_{k \in \mathcal{K}} \left[{R_{c,k}^u} \! -\! {{\log }_2}\left( {1 \!+\! {\Gamma _e}} \right)\right]}\tag{13a} \label{eq13a}\\
	&	{\rm{s.t.}} \kern 2pt\Pr \left( {\mathop {\max }\limits_{q \in \mathcal{Q}} \gamma _{p,q,k}^e \le {\Gamma _e}} \right) \ge \kappa, \forall k \in \mathcal{K}, \tag{13b} \label{eq13b}\\
	&
    \kern 15pt {\sum\limits_{i \in {{\cal K}_c}} {{\rm{Tr}}\left( {{{\bf{W}}_i}{{\bf{E}}^{(n)}}} \right)} \! + \!{\rm{Tr}}\left( {{\bf{B}}{{{\bf{\bar E}}}^{(n)}}} \right)\! \le \! {P_n}}, \forall n \in\mathcal{N}, \tag{13c} \label{eq13c}\\
	&
	\kern 17pt \gamma _{p,k}^u \ge {\Gamma  _{p,k}}, \forall k \in \mathcal{K}, \tag{13d} \label{eq13d}\\
	&
	\kern 17pt	{\rm{rank}}\left( {{{\bf{W}}_i}} \right) = 1,  \forall i \in {{\cal K}_c},  \tag{13e} \label{eq13e}
\end{align}
where ${{\bf{W}}_i} \buildrel \Delta \over = {{\bf{w}}_i}{\bf{w}}_i^H$, $i \in {{\cal K}_c}$, ${\bf{B}} \buildrel \Delta \over = {\bf{D}}{{\bf{D}}^H}$,  $\kappa$ is a probability factor for ensuring the security, and  ${{\bf{E}}^{(n)}}  \buildrel \Delta \over = {\bf{e}}_n{\bf{e}}_n^H$, ${{{\bf{\bar E}}}^{(n)}} \buildrel \Delta \over = {\bf{\bar e}}_n{\bf{\bar e}}_n^H$, with ${\bf{e}}_n \in {\mathbb{R}^{ N  \times 1}}$ denoting the $n$th unit vector, i.e., $[{\bf{e}}_n]_n=1, [{\bf{e}}_n]_i=0, \forall i \ne n$, ${\bf{\bar i}}_n={\bf{V}}_0^T{\bf{e}}_n$.  In constraint \eqref{eq13b}, we aim to limit the maximum received SINR among Eves to less than the SINR threshold ${\Gamma _e}$ with probability $\kappa$.

However, \textbf{P3} is still a non-convex program for the probabilistic/chance constraints, tightly coupled variables, and the rank-one constraints. Aiming at the above-mentioned difficulties, in the ;ing we first replace the constraint \eqref{eq13b} by a linear matrix inequality (LMI) constraint. As a compromise, we investigate a reformulation to serve as a lower bound for the original constraint as following lemma.

\textsl{Lemma 1:} The constraint \eqref{eq13b} is recast as
\begin{eqnarray}
	\setcounter{equation}{14}
	{{\bf{W}}_k} - {\Gamma _e}\sum\limits_{i \in {\mathcal{K}_c}\backslash k} {{{\bf{W}}_i} - {\Gamma _e}{{\bf{V}}_0}{\bf{B}}{\bf{V}}_0^H \preceq } {\bf{I}}_N\xi, \forall k \in \mathcal{K},
\label{eq14}\end{eqnarray}
where $\xi  = \Phi _N^{ - 1}\left( {1 - {\kappa ^{1/Q}}} \right){\Gamma _e}{\sigma _e^2}$, with $\Phi _N^{ - 1}\left(  \cdot  \right)$ indicating the inverse cumulative distribution function (c.d.f.) of an inverse central chi-square random variable with $2N$ DoF. 

\begin{proof}
	See Appendix A.
\end{proof}

\textsl{Remark 3:} According to the central limit theorem \cite{Fundamentals_Tse}, we assume that the Eve channels are modeled as Rayleigh fading channels since there are a large number of statistically independent reflected and scattered paths between the base station and the passive Eves. 

By replacing \eqref{eq13b} with \eqref{eq14}, \textbf{P3} can be reformulated as
\begin{align}
	\textbf{P4:} \kern 4pt & \mathop {\max }\limits_{{{\bf{W}}_i} \succeq \boldsymbol{0},{\bf{B}} \succeq \boldsymbol{0},{\Gamma _e}}  \kern 2pt  {\sum\limits_{k \in \mathcal{K}} \left[{R_{c,k}^u}  - {{\log }_2}\left( {1 + {\Gamma _e}} \right)\right]}\tag{15a} \label{eq15a}\\
	&{\rm{s.t.}} \kern 2pt{{\bf{W}}_k}\! -\! {\Gamma _e}\sum\limits_{i \in {\mathcal{K}_c}\backslash k} {{{\bf{W}}_i}\! -\! {\Gamma _e}{{\bf{V}}_0}{\bf{B}}{\bf{V}}_0^H \preceq } {\bf{I}}_N\xi, \forall k \in \mathcal{K}, \tag{15b} \label{eq15b}\\&
	\kern 15pt \eqref{eq13c},\eqref{eq13d}, \eqref{eq13e}\tag{15c}. \label{eq15c}
\end{align}

\textsl{Remark 4:} We would like to emphasize that the feasible solution of \textbf{P4} satisfies \textbf{P3} but not vice versa for the inequality transformation in \eqref{eq46}. 

Now, constraints ${\rm{rank}}\left( {{{\bf{W}}_i}} \right) = 1$, $ \forall i \in {{\cal K}_c}$ are still not convex, which are the remaining obstacle in solving \textbf{P4}. To make problem in a form suitable for semidefinite relaxation (SDR), we drop the rank constraints. The optimal matrix will usually not be rank-one for the rank relaxation. If it is, then its principal component is the optimal solution of the original problem. If not, then ${\rm{Tr}\left( {{{\bf{W}}_i}} \right)}$, $ \forall i \in {{\cal K}_c}$, is a lower bound needed to meet the constraints. Some ways of generating good solutions have been studied \cite{Multicasting_Luo}.

\subsection{Optimization Algorithm With a Fixed ${\Gamma _e}$} 
Intuitively, it is an equivalent solution with less difficulty by adjusting ${\Gamma _e}$. We can solve \textbf{P4} by first optimizing over $\left(\{{{\bf{W}}_i}\}_{i \in \mathcal{K}_c}, {\bf{B}}\right)$, and considering ${\Gamma _e}$ to be fixed. It is observed that the term ${{\log }_2}\left( {1 + {\Gamma _e}} \right)$ in objective can be dropped to simplify the optimization problem without affecting the optimal solution.

It can be easily found that \eqref{eq15a} is neither convex nor concave. To make \textbf{P4} a tractable problem, the non-concave parts are converted to corresponding lower bound function according to the following proposition \cite{Robust_Renhai}.

\textsl{Proposition 1:} Supposing a positive scalar $\chi_k$ and function $\mathscr{G}(\chi_k |\zeta_k ) = -\left( {\chi_k \zeta_k /{\rm{ln2}}} \right) + {{\rm{log}}_{\rm{2}}}\chi_k  + \left( {1/{\rm{ln2}}} \right)$, we have
\begin{eqnarray}
	\setcounter{equation}{16}	
	- {\log _2}\zeta_k = \mathop {\max }\limits_{\chi_k > 0} \kern 2pt \mathscr{G}(\chi_k|\zeta_k).
\label{eq16}\end{eqnarray}
The optimal solution of the right-hand side in \eqref{eq16} is $\chi_k = 1/\zeta_k$.

\begin{proof}
	Since $f(\chi )$ is concave, the optimal solution to the right-hand side is obtained when the gradient equals to 0, i.e., $\partial \mathscr{G}(\chi_k|\zeta_k)/\partial \chi_k=0$.
\end{proof}

By setting $\zeta_k  = \sum\limits_{i \in \mathcal{K}} {{\bf{h}}_{u,k}^H{{\bf{W}}_i}} {{\bf{h}}_{u,k}} + \sigma _u^2$, the corresponding surrogate lower bound functions, denoted by ${\mathscr{G}_{\zeta_k }}$, can be given by
\begin{align}
{\mathscr{G}_{\zeta_k }}\left(\chi_k \right)& = \mathscr{G}\left(\chi_k \left| \zeta_k=  \sum\limits_{i \in \mathcal{K}} {{\bf{h}}_{u,k}^H{{\bf{W}}_i}} {{\bf{h}}_{u,k}} + \sigma _u^2\right)  \right. \notag\\&
=  - \frac{\chi_k }{{\ln 2}}\left( {\sum\limits_{i \in \mathcal{K}} {{\bf{h}}_{u,k}^H{{\bf{W}}_i}} {{\bf{h}}_{u,k}} + \sigma _u^2} \right) + {\log _2}\chi_k  + \frac{1}{{\ln 2}}.
\label{eq17}\end{align}
Then, we obtain the affine function as
\begin{eqnarray}
	 - {\log _2}\left( {\sum\limits_{i \in \mathcal{K}} {{\bf{h}}_{u,k}^H{{\bf{W}}_i}} {{\bf{h}}_{u,k}} + \sigma _u^2} \right) = \mathop {\max }\limits_{\chi_k} \kern 2pt {\mathscr{G}_{{\zeta_k }}}(\chi_k ).
\label{eq18}\end{eqnarray}
To guarantee a tight approximation, in the $r$th iteration, $\chi_k^{(r)}$ is updated by
\begin{align}
{\chi_k ^{(r)}} = &\arg \mathop {\max }\limits_{\chi_k  > 0}  \kern 2pt - \frac{\chi_k }{{\ln 2}}\left( {\sum\limits_{i \in \mathcal{K}} {{\bf{h}}_{u,k}^H{\bf{W}}_i^{^{(r-1)}}} {{\bf{h}}_{u,k}} + \sigma _u^2} \right)  \notag \\
& + {\log _2}\chi_k + \frac{1}{{\ln 2}}.
\label{eq19}\end{align}
The closed-form solution of above problem is obtained by
\begin{eqnarray}
	{\chi_k ^{(r)}} = {\left( {\sum\limits_{i \in \mathcal{K}} {{\bf{h}}_{u,k}^H{\bf{W}}_i^{^{(r-1)}}} {{\bf{h}}_{u,k}} + \sigma _u^2} \right)^{ - 1}}.
	\label{eq20}\end{eqnarray}
After obtaining ${\chi_k ^{(r)}}$, \textbf{P4} is replaced by
\begin{align}
	&	\left\{ {{\bf{W}}_k^{(r)}} \right\}_{k \in \mathcal{K}_c} =  \notag\\&
	\arg \mathop {\max }\limits_{ {{{\bf{W}}_k}} } \sum\limits_{k \in {\cal K}} \left[{\mathscr{G}_{\zeta_k}}\left( {{\chi_k ^{(r)}}} \right) +  {{{\log }_2}\left( { \sum\limits_{i \in {\mathcal{K}}_c} {{\bf{h}}_{u,k}^H{{\bf{W}}_i}{{\bf{h}}_{u,k}}}  + \sigma _u^2} \right)} \right] \notag\\&
	{\rm{s.t.}} \kern 2pt \eqref{eq15b},\eqref{eq13c},\eqref{eq13d}.
	\label{eq21}\end{align}
Obviously, it is found that the problem \eqref{eq21} is a typical SDP problem. By means of standard convex software, such as CVX \cite{CVX_Grant} and SeDuMi \cite{SeDuMi_Sturm}, we can get the optimal solutions.
\begin{table}[t]
	\begin{center}
		\begin{tabular}{llr}
			\hline
			\textbf{Algorithm 1 }Joint design algorithm for \textbf{P4}\\
			\hline
			\textbf{Initialization:} Set $\{{\chi_k^{(0)}}\}_{k \in \mathcal{K}}: =1$, $\{{{\bf{W}}_k^{(r)}}\}_{k \in \mathcal{K}_c}:={\bf{I}}$, \\ 
		    \kern 10pt	$r:=0$, and tolerance $\epsilon_1>0$; \\
			1. \textbf{repeat}\\
			2. \kern 10pt	$r=r+1$;\\
			3. \kern 10pt 	Determine $(\{{{\bf{W}}_k^{(r)}}\}_{k \in \mathcal{K}_c},{\bf{B}})$ via CVX by substituting\\
			   \kern 20pt  ${\chi_k^{(r)}}$ into \eqref{eq21};\\
			4. \kern 10pt   Determine ${\chi_k^{(r)}}$ by substituting  $\{{{\bf{W}}_k^{(r)}}\}_{k \in \mathcal{K}}$ into \eqref{eq20}; \\
			5. \textbf{until} some convergence condition is met.\\
			\textbf{Output:} $(\{{{\bf{W}}_i^\star}\}_{i \in {\cal K}_c}, {\bf{B}}^\star)$. \\
			\hline
		\end{tabular}
		\label{tab1}
	\end{center}
\end{table}

The rudimentary procedure for \textbf{P4} is outlined in Algorithm 1. We prove that the Algorithm 1 theoretically converge to the optimal points presented in the following proposition.

\textsl{Proposition 2:} The sequence $( {\{ {{\bf{W}}_k^{(r)}} \}_{k \in \mathcal{K}},\{{\chi ^{(r)}_k}\}_{k \in \mathcal{K}}} )$ is generated by the Algorithm 1 with stationary convergence guarantee, which is a Karush-Kuhn-Tucker (KKT) point of the original problem \textbf{P4}.

\begin{proof}
	See Appendix B.
\end{proof}

\subsection{Optimization Over ${\Gamma _e}$}
Next, let us turn back to optimize over ${\Gamma _e}$ to obtain the global optimal solutions by one dimensional search. The F-SSR maximization problem over ${\Gamma _e}$ is then formulated as
\begin{eqnarray}
	\textbf{P5:} \kern 4pt \mathop {\max }\limits_{{\Gamma _e} > 0} \kern 2pt\mathscr{P} ({\Gamma _e}), \kern 4pt {\rm{s.t.}} \kern 2pt \eqref{eq15b},
	\label{eq22}\end{eqnarray}
where $\mathscr{P} \left({\Gamma _e}\right)$ is defined as
\begin{align}
&\mathscr{P} \left({\Gamma _e}\right) \buildrel \Delta \over =\mathop {\max }\limits_{{{\bf{W}}_i},{\bf{B}}} \kern 2pt \sum\limits_{k \in {\cal K}} \left[R_{c,k}^u - {{{\log }_2}\left( {1 + {\Gamma _{e}}} \right)}\right].
\label{eq23}\end{align}
Due to the concave objective function, we adopt an efficient method to find the global optimal ${\Gamma _e^\star}$. For this goal, we optimize over ${\Gamma _e}$ and keep $\left(\left\{{{\bf{W}}_i}\right\}_{k \in {\cal K}_c},{\bf{B}}\right)$ fixed. Then, we further simplify the optimization problem as
\begin{align}
\textbf{P6:}\kern 4pt	&\mathop {\min }\limits_{{\Gamma _{e}} > 0} \kern 2pt {{{\log }_2}\left( { {\Gamma _{e}}} \right)} \tag{24a} \label{eq24a}\\
& {\rm{s.t.}} \kern 2pt {{\boldsymbol{\Pi }}_k}{\Gamma _e} \succeq {{\bf{{W}}}_k}, \forall k \in \mathcal{K},\tag{24b} \label{eq24b}
\end{align}
where ${{\boldsymbol{\Pi }}_k}  \buildrel \Delta \over = {{\bf{I}}_N}\Phi _N^{ - 1}\left( {1 - {\kappa ^{1/Q}}} \right){\sigma _e^2}+{\sum _{i \in {{\cal K}_c}\backslash k}}{{\bf{W}}_i} + {{\bf{V}}_0}{\bf{B}}{\bf{V}}_0^H$. For the monotonicity of function ${\rm{log}_2}\left(  \cdot  \right)$, the objective is simplified to ${\log _2}\left( {1 + {\Gamma _e}} \right) \to {\log _2}\left( {{\Gamma _e}} \right)$.

The Lagrangian function of the reformulated problem, denoted by ${\mathcal{{\cal L}}_1}\left( {{\Gamma _e},{{\boldsymbol{\Psi }}_k}} \right)$, can be derived as
\begin{align}
\setcounter{equation}{24}
{\mathcal{{\cal L}}_1}\left( {{\Gamma _e},{{\boldsymbol{\Psi }}_k}} \right) = {\log _2}\left( {{\Gamma _e}} \right) + \sum\limits_{k \in {\cal K}} {{\rm{Tr}}} \left\{ {{{\boldsymbol{\Psi }}_k}\left( {{{\boldsymbol{\Pi }}_k}{\Gamma _e} - {{\bf{W}}_k}} \right)} \right\}.
\label{eq25}\end{align}
Actually, let us obtain the optimal solution of \textbf{P6} by solving the dual problem verified by the following theorem.

\textsl{Theorem 1:} The dual problem with implicit constraint ${ - \sum\limits_{k \in {\cal K}} {{\rm{Tr}}\left\{ {{{\boldsymbol{\Psi }}_k}{{\boldsymbol{\Pi }}_k}} \right\}} }>0$ is given by
\begin{align}
\mathop {\max }\limits_{\{{{\boldsymbol{\Psi }}_k}\}_{k \in {\cal K}}} \kern 2pt&{\mathcal D}_1({{\boldsymbol{\Psi }}_k}) =  - {\log _2}\left( { - \sum\limits_{k \in {\cal K}} {{\rm{Tr}}\left\{ {{{\boldsymbol{\Psi }}_k}{{\boldsymbol{\Pi }}_k}} \right\}} } \right)\notag\\& 
\kern 45pt	 - \sum\limits_{k \in {\cal K}} {{\rm{Tr}}} \left\{ {{{\boldsymbol{\Psi }}_k}{{\bf{W}}_k}} \right\} - 1.
\label{eq26}\end{align}

\begin{proof}
See Appendix C.
\end{proof}

An efficient method for solving the unconstrained problem \eqref{eq26} is BFGS method \cite{BFGS_Nezhad}, which exhibits a superior convergence rate to solve log concave function. Then, we get a simpler problem without affecting optimality as
\begin{align}
\mathop {\min }\limits_{\{{{\boldsymbol{\Psi }}_k}\}_{k \in {\cal K}}} \kern 2pt-{\mathcal D}_1({{\boldsymbol{\Psi }}_k}).
\label{eq27}\end{align}

The BFGS algorithm is summarized as

{\bf{Step1}}. Choose a initial point ${{\boldsymbol{\Psi }}_k^{(0)}}$, $r:=0$, and tolerance $\epsilon_2$;

{\bf{Step2}}. Update search direction: ${\rm{vec}}\{{\Delta {{\boldsymbol{\Psi }}_k}}\} ={\boldsymbol{X}}_k{\rm{vec}}\{{\nabla _{{{\boldsymbol{\Psi }}_k}}}{{\cal D}_1}({{\boldsymbol{\Psi }}_k^{(r)}})\}$, where vec$(\cdot)$ stacks columns of matrix into a single column vector, and  ${\nabla _{\bf{x}}}f\left(  \cdot  \right)$ denotes the gradient of $f\left(  \cdot  \right)$ with respect to ${\bf{x}}$;

{\bf{Step3}}. Obtain dual variables: ${{\boldsymbol{\Psi }}_k^{(r+1)}}: = {{\boldsymbol{\Psi }}_k^{(r)}} +  {\nu_k }\Delta {{\boldsymbol{\Psi }}_k},\forall k \in {\cal K}$, where ${\nu _k} = \arg \mathop {\min }\limits_{{\nu} \ge 0} \{ - {{{\cal D}}_1}({{\boldsymbol{\Psi }}_k^{(r)}} + \nu \Delta {{\boldsymbol{\Psi }}_k})\}$ indicates the step length;

{\bf{Step4}}. Calculate iterate variables: ${{\boldsymbol{\Xi }}_k}: =- {\nabla _{{{\boldsymbol{\Psi }}_k}}}{{\cal D}_1}( {{\boldsymbol{\Psi }}_k^{(r+1)}} )+ {\nabla _{{{\boldsymbol{\Psi }}_k}}}{{\cal D}_1}( {{{\boldsymbol{\Psi }}_k}^{(r)}})$, and ${\boldsymbol{\Lambda }}_k:={\nu _k}{\boldsymbol{X}}_k{\rm{vec}}\{{\nabla _{{{\boldsymbol{\Psi }}_k}}}{{\cal D}_2}({{\boldsymbol{\Psi }}_k^{(r)}})\}$;

{\bf{Step5}}. Update
\begin{align} 
{{\boldsymbol{X}}_k^{(r+1)}}: = &{{\boldsymbol{X}}_k^{(r)}} + \frac{{\left( {{\rm{Tr}}\left\{ {{{\boldsymbol{\Lambda }}_k}{{\boldsymbol{\Xi }}_k}} \right\} + {\rm{Tr}}\left\{ {{{\boldsymbol{R}}_1}{{\boldsymbol{X}}_k^{(r)}}} \right\}} \right){{\boldsymbol{R}}_4}}}{{{{\left( {{\rm{Tr}}\left\{ { {{\boldsymbol{\Lambda }}_k}{{\boldsymbol{\Xi }}_k}} \right\}} \right)}^2}}}\notag\\&
- \frac{{{{\boldsymbol{X}}_k^{(r)}}{{\boldsymbol{R}}_2} + {{\boldsymbol{R}}_3}{{\boldsymbol{X}}_k^{(r)}}}}{{{\rm{Tr}}\left\{ { {{\boldsymbol{\Lambda }}_k}{{\boldsymbol{\Xi }}_k}} \right\}}},
\label{eq28}\end{align}
where ${{\boldsymbol{R}}_1} = {\rm{vec}}\left( {{{\boldsymbol{\Xi }}_k}} \right){\rm{vec}}{\left( {{{\boldsymbol{\Xi }}_k}} \right)^T}$, ${{\boldsymbol{R}}_2} = {\rm{vec}}\left( {{{\boldsymbol{\Xi }}_k}} \right){\rm{vec}}{\left( { {{\boldsymbol{\Lambda }}_k}} \right)^T}$, ${{\boldsymbol{R}}_3} = {\rm{vec}}\left( { {{\boldsymbol{\Lambda }}_k}} \right){\rm{vec}}{\left( {{{\boldsymbol{\Xi }}_k}} \right)^T}$, and ${{\boldsymbol{R}}_4} = {\rm{vec}}\left( { {{\boldsymbol{\Lambda }}_k}} \right){\rm{vec}}{\left( { {{\boldsymbol{\Lambda }}_k}} \right)^T}$;

{\bf{Step6}}. If $\| {\sum\limits_{k \in  {\cal K}} {\boldsymbol{\Lambda }}_k } \|_2^2 < \epsilon_2 $, stop iteration;

{\bf{Step7}}. Otherwise, $r:=r+1$, return to Step 2.

\begin{table}[t]
\begin{center}
	\begin{tabular}{llr}
		\hline
		\textbf{Algorithm 2 }Two-stage iterative algorithm for \textbf{P1}\\
		\hline
		\textbf{Initialization:} Pick up ${{\boldsymbol{\Psi }}_k}:={\bf{I}}, {{\boldsymbol{\Xi }}_k}:={\boldsymbol{0}}, {{\boldsymbol{X}}_k}: = {\bf{I}}$, and \\ \kern 10pt tolerance $\epsilon_2>0$; \\
		1. \textbf{repeat}\\
		2. \kern 10pt Compute $(\left\{{{\bf{W}}_i}\right\}_{i \in {\cal K}_c}, {\bf{B}} )$ in \eqref{eq21} as Algorithm 1;\\
		3. \kern 10pt 	Determine $\{{\boldsymbol{\Pi }}_k\}_{k \in {{\cal K}}}$ by $(\left\{{{\bf{W}}_i}\right\}_{i \in {\cal K}_c}, {\bf{B}} )$;\\
		4. \kern 10pt   Update ${{\boldsymbol{\Psi }}_k}$, ${{\boldsymbol{\Xi }}_k}$, and ${{\boldsymbol{X}}_k}$ as the BFGS algorithm; \\
		5. \kern 10pt   Determine ${{\Gamma _e}}$ by substituting ${{\boldsymbol{\Psi }}_k}$, and ${{\boldsymbol{\Xi }}_k}$ into \eqref{eq29}; \\
		6. \textbf{until} $\left\| {{{\bf{\Xi }}_k}} \right\|_2^2 <\epsilon_2$; \\
		\textbf{Output:} $(\{{{\bf{W}}_i^\star}\}_{i \in {\cal K}_c}, {\bf{B}^\star}, {{\Gamma _e^\star}})$. \\
		\hline
	\end{tabular}
	\label{tab3}
\end{center}
\end{table}

\textsl{Corollary 1:} The closed-form solution of \textbf{P6} can be got from that of \eqref{eq27} as
\begin{align}
\Gamma _e^\star =  - \frac{1}{{\sum\limits_{k \in {\cal K}}  {\rm{Tr}} \left({{\boldsymbol{\Psi }}_k} {{{\boldsymbol{\Pi }}_k}} \right)}}.
\label{eq29}\end{align}

\begin{proof}
After obtaining the optimal dual variables ${{\boldsymbol{\Psi }}_k^\star}$ and ${{\boldsymbol{\Xi }}_k^\star}$, \textbf{P6} boils down to find the optimal ${\Gamma _e}$ such that the KKT condition of \eqref{eq25} is satisfied, i.e.,
\begin{align}
	&{\nabla _{{\Gamma _e}}}{{\cal L}_1}\left( {{\Gamma _e},{{\boldsymbol{\Psi }}_k}} \right) = 0\notag\\&
	\Leftrightarrow \frac{1}{{{\Gamma _e}}} + \sum\limits_{k \in {\cal K}} {\rm{Tr}} \left({{\boldsymbol{\Psi }}_k} {{{\boldsymbol{\Pi }}_k}} \right) = 0.
	\label{eq30}\end{align}
Thus, Corollary 1 follows.
\end{proof}

Finally, we summarize the two-stage transmit information and AN beamforming design algorithm for the F-SSR maximization problem in Algorithm 2. Specifically, we first get the optimal beamforming vectors and AN under fixed allowable SINR; Next, based on the optimal values, we search the optimal allowable SINR. 
\section{Efficient Algorithm Design}
Solving a convex SDP via the interior-point method directly is inefficient due to a Hermitian matrix of $N^2$ variables in particular with a large-scale transmit array. By introducing an auxiliary variable ${\Sigma _p}$ following the spirit of \cite{Joint_Nguyen}, \textbf{P4} is transformed as
\begin{align}
	\textbf{P7:} \kern 4pt& \mathop {\max }\limits_{{{\bf{W}}_i\succeq \boldsymbol{0}},{\bf{B}}\succeq \boldsymbol{0}} \kern 2pt  \sum\limits_{k \in \mathcal{K}} {{{\log }_2}\left( {1 + {\bf{h}}_{u,k}^H{{\bf{W}}_c}{{\bf{h}}_{u,k}}} \right)}\tag{31a} \label{eq31a}\\
	&{\rm{s.t.}} \kern 2pt{{\bf{W}}_k}\! -\! {\Gamma _e}\sum\limits_{i \in {\mathcal{K}_c}\backslash k} {{{\bf{W}}_i}\! -\! {\Gamma _e}{{\bf{V}}_0}{\bf{B}}{\bf{V}}_0^H \preceq } {\bf{I}}_N\xi, \forall k \in \mathcal{K}, \tag{31b} \label{eq31b}\\
	&
	\kern 15pt {\sum\limits_{i \in {{\cal K}_c}} {{\rm{Tr}}\left( {{{\bf{W}}_i}{{\bf{E}}^{(n)}}} \right)} \! + \!{\rm{Tr}}\left( {{\bf{B}}{{{\bf{\bar E}}}^{(n)}}} \right)\! \le \! {P_n}},  \forall n \in\mathcal{N}, \tag{31c} \label{eq31c}\\
	&
	\kern 15pt	\gamma _{p,k}^u \ge {\Gamma  _{p,k}}, \forall k \in \mathcal{K}, \tag{31d} \label{eq31d}\\
	&
	\kern 15pt	\sum\limits_{i \in \mathcal{K}} {{\bf{h}}_{u,k}^H{{\bf{W}}_i}{{\bf{h}}_{u,k}}} + \sigma _u^2 \le {\Sigma _p}, \forall k \in \mathcal{K}. \tag{31e} \label{eq31e}
\end{align}

\textbf{P7} can be reforged to a more amenable form by utilizing the special features. We can solve the problem via a simple form as following theorem. 

\textsl{Theorem 2:} \textbf{P7} can be equivalently formulated as following mini-max problem
\begin{align}
	\textbf{P8:} \kern 4pt&\mathop {\min }\limits_{{{\bf{\tilde D}}_k} \succeq  \boldsymbol{0},{\boldsymbol{\tilde \eta }} \ge 0}\kern 4pt \mathop {\max }\limits_{\tilde \omega  \ge 0,{{\bf{\tilde W}}_k} \succeq  \boldsymbol{0},{\bf{\tilde B}} \succeq  \boldsymbol{0}} \sum\limits_{k \in  {\cal K}} {{{\log }_2}\frac{{\left| {{\bf{\tilde \Sigma }} + {{\bf{h}}_{u,k}}\tilde \omega {\bf{h}}_{u,k}^H} \right|}}{{\left| {{\bf{\tilde \Sigma }}} \right|}}} \tag{32a} \label{eq32a}\\&
	{\rm{s.t.}}\kern 4pt\tilde \omega  + \sum\limits_{k \in {\cal K}} {{\rm{Tr}}\left( {{{\bf{\Omega }}_k}{{{\bf{\tilde W}}}_k}} \right)}  + {\rm{Tr}}\left( {{\bf{\Phi \tilde B}}} \right) \le \vartheta,\tag{32b} \label{eq32b}\\&
	\kern 17pt \xi \sum\limits_{k \in {\cal K}} {{\rm{Tr}}\left( {{{{\bf{\tilde D}}}_k}} \right)}  + {{\bf{g}}^T}{\boldsymbol{\tilde \eta }} \le \vartheta,\tag{32c} \label{eq32c}
\end{align}
where ${\bf{\Sigma }} =  {\rm{diag}}({\boldsymbol{\lambda }}) - \sum\limits_{k \in {\cal K}} {{\Gamma _e}{{\bf{D}}_k}}$, ${\bf{\Phi }} = \sum\limits_{n \in \cal N} {\lambda _n}{{{\bf{\bar E}}}^{(n)}} - \sum\limits_{k \in {\cal K}} {{\Gamma _e}{{\bf{V}}_0}{{\bf{D}}_k}{\bf{V}}_0^H}$, ${{\bf{\Omega }}_k} = {{\bf{D}}_k} +  {\rm{diag}}({\boldsymbol{\lambda }}) - {\mu _k}{{\bf{H}}_{u,k}} + {\upsilon _k}{{\bf{H}}_{u,k}} + \sum\limits_{i \in {\cal K}\backslash k} {\left( {{\mu _i}{\Gamma _{p,i}}{{\bf{H}}_{u,i}} + {\upsilon _i}{{\bf{H}}_{u,i}} - {\Gamma _e}{{\bf{D}}_i}} \right)}$,  and ${\boldsymbol{\eta }} = {\left[ {{{\boldsymbol{\lambda }}^T},{{\boldsymbol{\mu }}^T},{{\boldsymbol{\upsilon }}^T}} \right]^T}$. Then, we have
\begin{align}
\setcounter{equation}{32}
	{{{\bf{W}}}_c} = \frac{1}{{\left\| {{{\bf{\Sigma }}^{ - 1/2}}{{\bf{h}}_{u,k}}} \right\|_2^2}}{{\bf{\Sigma }}^{ - 1}}{{\bf{h}}_{u,k}}\omega {\bf{h}}_{u,k}^H{{\bf{\Sigma }}^{ - 1}}.
\label{eq33}\end{align}

\begin{proof}
	 See Appendix D.
\end{proof}

\textsl{Remark 5:} One can verify that the inequality constraints in  \eqref{eq32b} and \eqref{eq32c} hold with equality at the optimum. As proof, assuming $\tilde \omega  + \sum\limits_{k \in {\cal K}} {{\rm{Tr}}\left( {{{\bf{\Omega }}_k}{{{\bf{\tilde W}}}_k}} \right)}  + {\rm{Tr}}\left( {{\bf{\Phi \tilde B}}} \right) < \vartheta$ for a given $\left({{{\bf{\tilde D}}_k}, {\boldsymbol{\tilde \eta }}}\right)$, there exists an arbitrarily small number $\delta>0$ satisfying $\tilde \omega  + \sum\limits_{k \in {\cal K}} {{\rm{Tr}}\left( {{{\bf{\Omega }}_k}{{{\bf{\tilde W}}}_k}} \right)}  + {\rm{Tr}}\left( {{\bf{\Phi \tilde B}}} \right)+ \delta < \vartheta$. Let’s replace $\tilde \omega$ by the objective function $\tilde \omega+\delta/2$ so as to yield a larger objective value. This creates contradiction due to the optimal value $\tilde \omega$. The same conclusion can be drawn in constraint $\xi \sum\limits_{k \in {\cal K}} {{\rm{Tr}}\left( {{{{\bf{\tilde D}}}_k}} \right)}  + {{\bf{g}}^T}{\boldsymbol{\tilde \eta }} < \vartheta$ for a given $\left(\tilde \omega,{{\bf{\tilde W}}_k},{\bf{\tilde B}}\right)$. This way, it facilitates a barrier interior point method to solve \textbf{P8} since equality constraints are usually easier to get access. Interestingly, solving \textbf{P8} with respect to $\tilde \omega$ demands for much lower complexity compared with optimizing ${{\bf{W}}_c}$. 

For ease of exposition, we first define
\begin{align}
&{{{\bf{\tilde W}}}_k}  \buildrel \Delta \over =  {\bf{\Omega }}_k^{^{ - 1/2}}{{{\bf{\hat W}}}_k}{\bf{\Omega }}_k^{^{ - 1/2}},\label{eq34} \\&
{\bf{\tilde B}} \buildrel \Delta \over = {{\bf{\Phi }}^{ - 1/2}}{\bf{\hat B}}{{\bf{\Phi }}^{ - 1/2}}.
	\label{eq35}\end{align}

As a standard step, the modified objective, denoted by $\mathscr{B}({\boldsymbol{\tilde \eta }},{{{\bf{\tilde D}}}_k},\tilde \omega ,{{{\bf{\hat W}}}_k},{\bf{\hat B}},\varsigma)$, can be reorganized as
\begin{align}
\setcounter{equation}{35}
&\mathscr{B}({\boldsymbol{\tilde \eta }},{{{\bf{\tilde D}}}_k},\tilde \omega ,{{{\bf{\hat W}}}_k},{\bf{\hat B}},\varsigma ) = 	\sum\limits_{k \in {\cal K}} {{{\log }_2}\frac{{\left| {{\bf{\tilde \Sigma }} + {{\bf{h}}_{u,k}}\tilde \omega {\bf{h}}_{u,k}^H} \right|}}{{\left| {{\bf{\tilde \Sigma }}} \right|}}} \notag \\& 
\kern 20pt + \frac{1}{\varsigma }\left\{ {{{\log }_2} (\tilde \omega ) + \sum\limits_{k \in {\cal K}} {{{\log }_2}\left[ {{\rm{Tr}}\left( {{{{\bf{\hat W}}}_k}} \right)} \right]} } + {{\log }_2}\left[ {{\rm{Tr}}\left( {{\bf{\hat B}}} \right)} \right] \right.\notag \\&
\kern 20pt 	\left. { - \sum\limits_{k \in {\cal K}} {{{\log }_2}\left[ {{\rm{Tr}}\left( {{{{\bf{\tilde D}}}_k}} \right)} \right]}  - \sum {{{\log }_2} ({{\tilde \eta }_i})} } \right\},
\label{eq36}\end{align}
where \eqref{eq36} is equivalent to problem \textbf{P8} when $\varsigma  \to \infty $, ${\log (\tilde \omega )}$, ${\log ({{\tilde \eta }_i})}$, ${{{\log }_2}[ {{\rm{Tr}}( {{{{\bf{\tilde D}}}_k}} )} ]}$, ${{{\log }_2}[ {{\rm{Tr}}( {{\bf{\hat B}}} )} ]}$, and ${{{\log }_2}[ {{\rm{Tr}}( {{{{\bf{\hat W}}}_k}} )} ]}$ denote the logarithmic barrier factors to explicate the non-negativity constraints, ${\tilde \omega>0}$, ${{\tilde \eta }_i}>0$, ${\rm{Tr}}( {{{{\bf{\tilde D}}}_k}} )>0$, ${{\rm{Tr}}( {{\bf{\hat B}}} )}>0$, and ${\rm{Tr}}( {{{{\bf{\hat W}}}_k}} )>0$, respectively, and $\varsigma$ denotes regulation term for the logarithm barrier terms. A standard equality constrained problem can be solved with a fixed $\varsigma$, i.e., 
\begin{align}
	\textbf{P9:} \kern 4pt&\mathop {\min }\limits_{{{\bf{\tilde D}}_k} \succeq  \boldsymbol{0},{\boldsymbol{\tilde \eta }} \ge 0}\kern 4pt \mathop {\max }\limits_{\tilde \omega  \ge 0,{{\bf{\hat W}}_k} \succeq  \boldsymbol{0},{\bf{\hat B}} \succeq  \boldsymbol{0}} \sum\limits_{k \in  {\cal K}} {\mathscr{B}({\boldsymbol{\tilde \eta }},{{{\bf{\tilde D}}}_k},\tilde \omega ,{{{\bf{\hat W}}}_k},{\bf{\hat B}},\varsigma ) } \tag{37a} \label{eq37a}\\&
	{\rm{s.t.}}\kern 4pt\tilde \omega  + \sum\limits_{k \in {\cal K}} {{\rm{Tr}}\left( {{{\bf{\hat W}}}_k} \right)}  + {\rm{Tr}}\left( {\bf{\hat B}} \right) = \vartheta,\tag{37b} \label{eq37b}\\&
	\kern 17pt \xi \sum\limits_{k \in {\cal K}} {{\rm{Tr}}\left( {{{{\bf{\tilde D}}}_k}} \right)}  + {{\bf{g}}^T}{\boldsymbol{\tilde \eta }} = \vartheta.\tag{37c} \label{eq37c}
\end{align}
The procedure of the barrier method is to search the optimal solutions with a fixed $\varsigma $, and then adjust $\varsigma $ until some stopping criterion is satisfied. It is widely known that the infeasible-start Newton’s method for solving a mini-max optimization problem shows a faster rate of convergence. We start with the necessary and sufficient optimal conditions (i.e. the KKT conditions  \cite{Dirty_Tran}) for \textbf{P9} as
\begin{align}
{\nabla _{\tilde \omega }}{\mathscr{B}}({\boldsymbol{\tilde \eta }},{{{\bf{\tilde D}}}_k},\tilde \omega ,{{{\bf{\hat W}}}_k},{\bf{\hat B}},\varsigma ) - {\tau _1} = 0,  \tag{38a} \label{eq40a}
\end{align}
\begin{align}
{\nabla _{{{{\bf{\hat W}}}_k}}}{\mathscr{B}}({\boldsymbol{\tilde \eta }},{{{\bf{\tilde D}}}_k},\tilde \omega ,{{{\bf{\hat W}}}_k},{\bf{\hat B}},\varsigma ) - {\tau _1}{{\bf{I}}_N} = 0,\tag{38b} \label{eq38b}
\end{align}
\begin{align}
{\nabla _{{\bf{\hat B}}}}{\mathscr{B}}({\boldsymbol{\tilde \eta }},{{{\bf{\tilde D}}}_k},\tilde \omega ,{{{\bf{\hat W}}}_k},{\bf{\hat B}},\varsigma ) - {\tau _1}{{\bf{I}}_N} = 0,,\tag{38c} \label{eq40c}
\end{align}
\begin{align}
{\nabla _{{{{\bf{\tilde D}}}_k}}}{\mathscr{B}}({\boldsymbol{\tilde \eta }},{{{\bf{\tilde D}}}_k},\tilde \omega ,{{{\bf{\hat W}}}_k},{\bf{\hat B}},\varsigma ) + {\tau _2}\xi {{\bf{I}}_N} = 0,\tag{38d} \label{eq38d}
\end{align}
\begin{align}
{\nabla _{\boldsymbol{\lambda }}}{\mathscr{B}}({\boldsymbol{\tilde \eta }},{{{\bf{\tilde D}}}_k},\tilde \omega ,{{{\bf{\hat W}}}_k},{\bf{\hat B}},\varsigma ) + {\tau _2}{{\bf{g}}^T}{{\bf{I}}_N} = 0,\tag{38e} \label{eq38e}
\end{align}
\begin{align}
{\nabla _{\boldsymbol{\mu }}}{\mathscr{B}}({\boldsymbol{\tilde \eta }},{{{\bf{\tilde D}}}_k},\tilde \omega ,{{{\bf{\hat W}}}_k},{\bf{\hat B}},\varsigma ) + {\tau _2}{{\bf{g}}^T}{{\bf{I}}_N} = 0,\tag{38f} \label{eq38f}
\end{align}
\begin{align}
{\nabla _{\boldsymbol{\upsilon }}}{\mathscr{B}}({\boldsymbol{\tilde \eta }},{{{\bf{\tilde D}}}_k},\tilde \omega ,{{{\bf{\hat W}}}_k},{\bf{\hat B}},\varsigma ) + {\tau _2}{{\bf{g}}^T}{{\bf{I}}_N} = 0,\tag{38g} \label{eq38g}
\end{align}
\begin{align}
\tilde \omega  + \sum\limits_{k \in {\cal K}} {{\rm{Tr}}\left( {{{{\bf{\hat W}}}_k}} \right)}  + {\rm{Tr}}\left( {{\bf{\hat B}}} \right) = \vartheta, \tag{38h} \label{eq38h}
\end{align}
\begin{align}
\xi \sum\limits_{k \in {\cal K}} {{\rm{Tr}}\left( {{{{\bf{\tilde D}}}_k}} \right)}  + {{\bf{g}}^T}{\boldsymbol{\tilde \eta }} = \vartheta. \tag{38l} \label{eq38l}
\end{align}
In particular, the Newton step $( {\Delta \tilde \omega ,\Delta {{{\bf{\hat W}}}_k},\Delta {\bf{\hat B}},\Delta {\boldsymbol{\tilde \eta }},\Delta {{{\bf{\tilde D}}}_k}} )$ is computed for updating $( {\tilde \omega ,{{{\bf{\hat W}}}_k},{\bf{\hat B}},{\boldsymbol{\tilde \eta }},{{{\bf{\tilde D}}}_k}} )$ in \eqref{eq40a} as
\begin{align}
	\setcounter{equation}{38}
&\varsigma\sum\limits_{k \in {\cal K}} {{\bf{h}}_{u,k}^H{{\left( {{\bf{\tilde \Sigma }} + {{\bf{h}}_{u,k}}\tilde \omega {\bf{h}}_{u,k}^H + \Delta {\bf{\tilde \Sigma }} + {{\bf{h}}_{u,k}}\Delta \tilde \omega {\bf{h}}_{u,k}^H} \right)}^{ - 1}}{{\bf{h}}_{u,k}}} \notag\\&
 + {\left( {\tilde \omega  + \Delta \tilde \omega } \right)^{ - 1}} - \varsigma \left( {{\tau _1} + \Delta {\tau _1}} \right) = 0,
\label{eq39}\end{align}
where $\Delta {\bf{\tilde\Sigma }} = {\rm{diag}}(\Delta {\boldsymbol{\lambda }}) - \sum\limits_{k \in {\cal K}} {{\Gamma _e}\Delta {{\bf{D}}_k}}$. Applying matrix inverse lemma, $\footnote{{The  matrix inverse approximation $({\bf{X}} + {\bf{Y}})^{-1} \simeq {{\bf{X}}^{ - 1}} - {{\bf{X}}^{ - 1}}{\bf{Y}}{{\bf{X}}^{ - 1}}$ can be applied to small entries of matrix ${\bf{Y}}$ \cite[Ch. 3]{Matrix_Meyer}. When Algorithm 3 closes to optimal value as the iterative number increasing, the residual error can be ignored.}}$ then, \eqref{eq39} is approximated by
\begin{align}
&\sum\limits_{k \in  {\cal K}} {\left( {\varsigma {{\tilde \omega }^2}{\bf{h}}_{u,k}^H{{\bf{F}}_k}\Delta {\bf{\tilde \Sigma }}{{\bf{F}}_k}{{\bf{h}}_{u,k}}\! +\! \varsigma {{\tilde \omega }^2}{\bf{h}}_{u,k}^H{{\bf{F}}_k}{{\bf{h}}_{u,k}}\Delta \tilde \omega {\bf{h}}_{u,k}^H{{\bf{F}}_k}{{\bf{h}}_{u,k}}} \right)}\notag \\&
 + \Delta \tilde \omega  + \varsigma {{\tilde \omega }^2}\Delta {\tau _1} = \sum\limits_{k \in  {\cal K}} {\varsigma {{\tilde \omega }^2}{\bf{h}}_{u,k}^H{{\bf{F}}_k}{{\bf{h}}_{u,k}}}  + \tilde \omega  - \varsigma {{\tilde \omega }^2}{\tau _1},
 \label{eq40}\end{align}
where ${{\bf{F}}_k} = {\left( {{\bf{\tilde \Sigma }} + {{\bf{h}}_{u,k}}\tilde \omega {\bf{h}}_{u,k}^H} \right)^{ - 1}}$. 

Then, we use each Newton step to stack as a system of linear equations \cite{Convex_Boyd}.

The rudimentary numerical algorithm to obtain the optimal solutions of problem \textbf{P9} is outlined in Algorithm 3.
\begin{table}[t]
	\begin{center}
		\begin{tabular}{llr}
			\hline
			\textbf{Algorithm 3 }Barrier algorithm for \textbf{P9}\\
			\hline
			\textbf{Initialization:} Pick up ${{\bf{\hat W}}_k}: = {{\bf{I}}_N},{\bf{\hat B}}: = {{\bf{I}}_{N - K}},{{\bf{\tilde D}}_k}: = {{\bf{I}}_N}$,\\
			 \kern 10pt $\tilde \omega : = 1, {\boldsymbol{\tau }}: = {\boldsymbol{0}}, \varsigma : = {\varsigma _0},{\boldsymbol{\lambda }}: = {\boldsymbol{1}},{\boldsymbol{\mu }}: = {\boldsymbol{1}},{\boldsymbol{\upsilon }}: = {\boldsymbol{1}}$, $\ell$, and tolerance\\
			 \kern 10pt $\epsilon_3>0$; \\
			1. \textbf{repeat}\{Outer iteration\}\\
			2. \kern 5pt  \textbf{repeat} \{Inner iteration (centering step)\}\\
			3. \kern 10pt  Obtain the Newton step $( {\Delta \tilde \omega ,\Delta {{{\bf{\hat W}}}_k},\Delta {\bf{\hat B}},\Delta {\boldsymbol{\tilde \eta }},\Delta {{{\bf{\tilde D}}}_k}}, \Delta {\boldsymbol{\tau }})$ \\
			\kern 20pt from linear equations;\\
			4. \kern 10pt  Backtracking line search:\\
			5. \kern 15pt  $s = 1$;\\
			6. \kern 15pt \textbf{while}\kern 2pt$r( {\tilde \omega  + s\Delta \tilde \omega ,{{{\bf{\hat W}}}_k} + s\Delta {{{\bf{\hat W}}}_k},{\bf{\hat B}} + s\Delta {\bf{\hat B}},{\boldsymbol{\eta }} + s\Delta {\boldsymbol{\eta }}}, {{{\bf{\tilde D}}}_k}$\\
			\kern 24pt $ + s\Delta {{{\bf{\tilde D}}}_k},{\boldsymbol{\tau }} + s\Delta {\boldsymbol{\tau }}) > ( {1 - \alpha s} )r( \tilde \omega ,{{{\bf{\hat W}}}_k},{\bf{\hat B}},{\boldsymbol{\eta }},{{{\bf{\tilde D}}}_k},{\boldsymbol{\tau }} );$ \textbf{do}\\
			7. \kern 15pt  $s = \beta s;$\\
			8. \kern 15pt \textbf{end while}\\
			9. \kern 10pt Compute primal and dual variables: ${\tilde \omega}:={\tilde \omega}  + s\Delta \tilde \omega, {{{\bf{\hat W}}}_k}:=$\\
			   \kern 20pt ${{{{\bf{\hat W}}}_k} + s\Delta {{{\bf{\hat W}}}_k},{\bf{\hat B}}:= {\bf{\hat B}} + s\Delta {\bf{\hat B}}, {\boldsymbol{\eta }}:={\boldsymbol{\eta }} + s\Delta {\boldsymbol{\eta }}},{{{\bf{\tilde D}}}_k}:={{{\bf{\tilde D}}}_k}$\\
			   \kern 20pt $+s\Delta {{{\bf{\tilde D}}}_k},  {\boldsymbol{\tau }}:={\boldsymbol{\tau }} + s\Delta {\boldsymbol{\tau }};$\\
			10.\kern 5pt \textbf{until} $r( \tilde \omega ,{{{\bf{\hat W}}}_k},{\bf{\hat B}},{\boldsymbol{\eta }},{{{\bf{\tilde D}}}_k},{\boldsymbol{\tau }} )<\epsilon_3;$\\
			11.\kern 5pt Increase $\varsigma:=\ell \varsigma$;\\
			12.\textbf{until} $t$ is sufficiently large to tolerate the duality gap. \\
			\hline
		\end{tabular}
		\label{tab2}
	\end{center}
\end{table}

\section{Simulation Results}
In this section, we highlight the advantages of the proposed schemes by comparing the secure performance with other reference schemes through numerical simulation. The carrier frequency is chosen ${f_c} = 1$ {\rm{GHz}}. The IoDs' number is set as $K=2$ with directions ${\theta _{u,1}}=-35^\circ$, ${\theta _{u,2}}=15^\circ$, respectively. To highlight the performance gain contributed from beamforming, the reference distance is set to 1000 meters for all IoDs. For simplicity, the per-antenna power constraint is $P_n = P_\text{Tol}/N$, $\forall n\in \mathcal{N}$. The background thermal noise variances are $\sigma _{u}^2=\sigma _{u,k}^2=-100$ dBm, $\forall k \in \mathcal{K}$, and $\sigma _{e}^2=\sigma _{e,q}^2=-100$ dBm, $\forall q \in \mathcal{Q}$. The minimum desired received SINR of the private stream is identical for each IoD, i.e., ${\Gamma  _{p}}={\Gamma  _{p,k}}$, $\forall k  \in \mathcal{K}$. The number of Eves is $Q=2$. The probability parameter is picked up $\kappa=0.95$. Two existing baseline schemes, i.e., secrecy rate maximization (SRM) \cite{Robust_Fu} and maximum ratio transmission (MRT) \cite{CSIT_Joudeh}, are used to compare with our proposed scheme. The signal attenuation factor, denoted by $\rho \left( r \right)$, is determined by \cite{Wireless_Goldsmith}
\begin{eqnarray}
	\setcounter{equation}{41}
	\begin{aligned}[b]
		{\rm{Lfs}}(\text{dB}) &= - 20{\log }[\rho (r)]\\&
		= 32.5 + 20\log[{f_c}(\text{MHz})] + 20\log[r(\text{Km})].
	\end{aligned}
	\label{eq41}\end{eqnarray}

\begin{figure}[tb]
	\begin{center}
		\includegraphics[width=1\columnwidth]{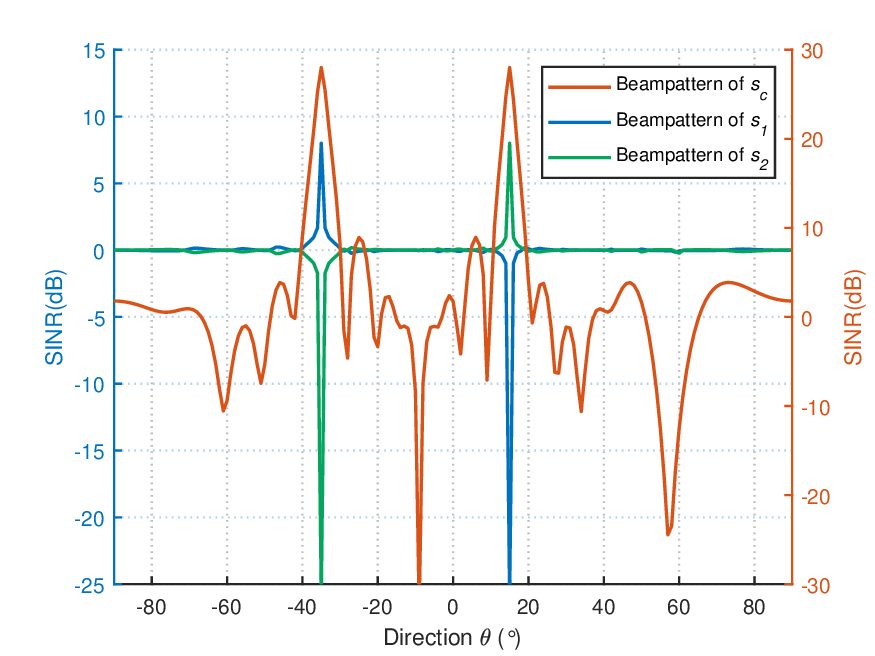}
	\end{center}
	\caption{The transmit common and private stream beampatterns versus direction for the proposed method, where $N=20$, ${\Gamma_{p}}=8$ dB, and ${P_\text{Tol}}=10$ dBm.}
	\label{fig2}
\end{figure}

According to the analysis in Section II, we know that the multiple private streams need to simultaneously transmit toward corresponding IoDs and the common streams need to transmit toward all IoDs. For the case with $N=20$, ${\Gamma_{p}}=8$ dB, and ${P_\text{Tol}}=10$ dBm. We explore the transmit beampatterns versus direction in Fig. \ref{fig2}, where the received SINR values of the common and private streams correspond to right-hand axis and left-hand axis, respectively. As expected, two sharp SINR peaks for the common streams are formed in the directions of IoDs. The SINRs of private streams along the directions of corresponding IoDs are fully compliant with the predefined requirements whose values are poised above the required SINR 8 dB. Those guarantee reliable transmissions of the private streams from the based station to IoDs. One can see that the private streams are hidden deep in the common streams, intuitively. In contrast, the performed SINRs are so poor in undesired directions.

\begin{figure}[tb]
	\begin{center}
		\includegraphics[width=1\columnwidth]{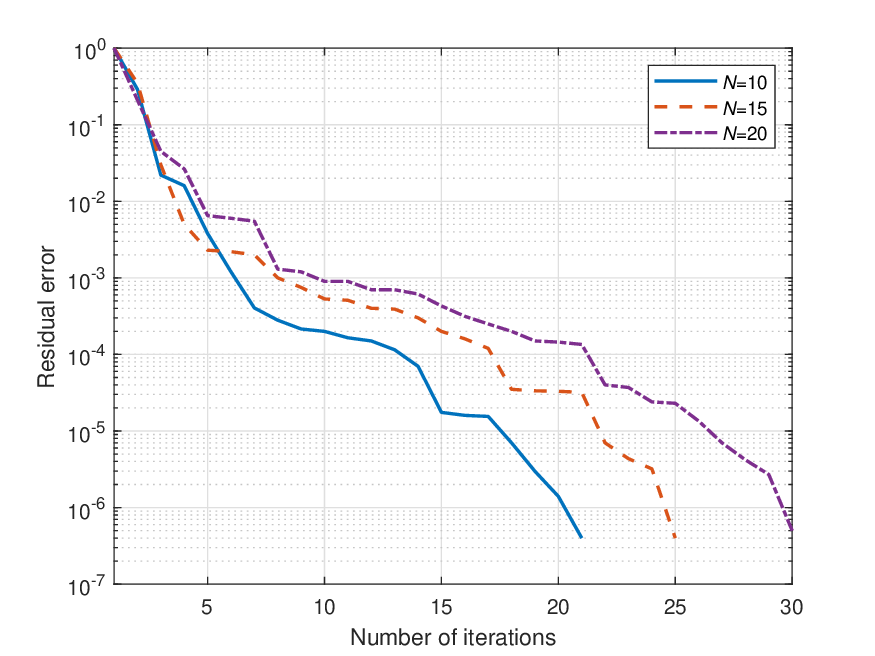}
	\end{center}
	\caption{Convergence rate of the proposed Algorithm 1 for different number of transmit antennas, where ${\Gamma  _{p}}=8$ dB, $\epsilon_1=10^{-6}$, and ${P_\text{Tol}}=10$ dBm.}
	\label{fig3}
\end{figure}

Figure \ref{fig3} shows the convergence rate of Algorithm 1 with iteration initial values starting from random points. We can observe that Algorithm 1 converges faster, as well as it is slightly sensitive to the system configurations.

\begin{figure}[tb]
	\begin{center}
		\includegraphics[width=1\columnwidth]{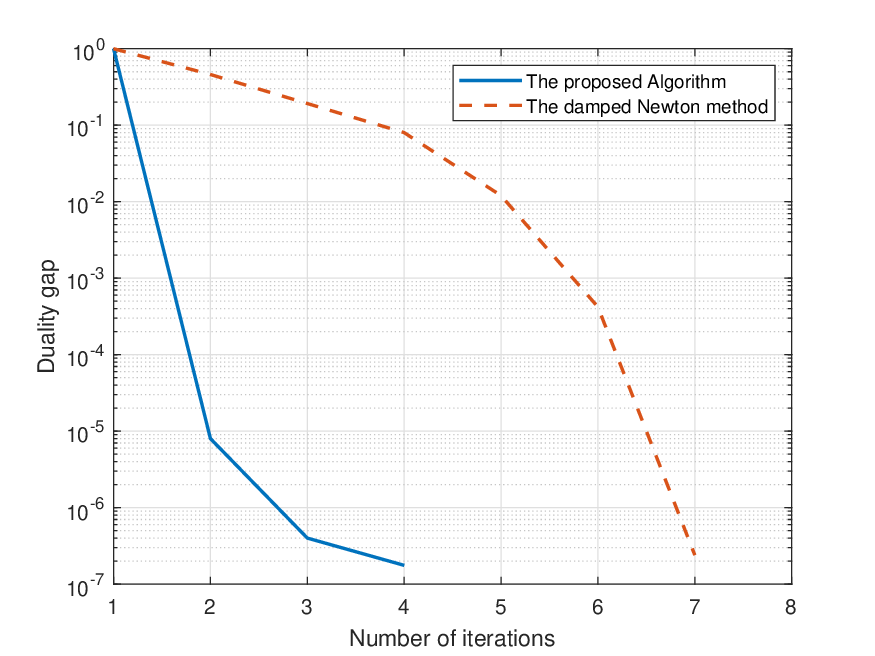}
	\end{center}
	\caption{Convergence rate of the proposed Algorithm 2, where ${\Gamma  _{p}}=8$ dB, $N=20$, $\epsilon_2=10^{-6}$, and ${P_\text{Tol}}=10$ dBm.}
	\label{fig4}
\end{figure}

In Fig. \ref{fig4} , we depict our proposed  Algorithm 2 in terms of convergence behavior, and we also compare with the damped Newton method. Admittedly, one way to solve the considered problem via the damped Newton method follows similar steps as in \cite[Sec. 9.5.2]{CVX_Grant}. The initial value of the maximum allowable SINR of the private stream for Eves is set as 0 dB. It is found that the Algorithm 2 can obtain the same duality gap as damped Newton method with fewer iterations. This is expected since the BFGS algorithm is very effective self-correcting properties in dealing with the inverse of the true Hessian matrix \cite{BFGS_Nezhad}.

\begin{figure}[tb]
	\begin{center}
		\includegraphics[width=1\columnwidth]{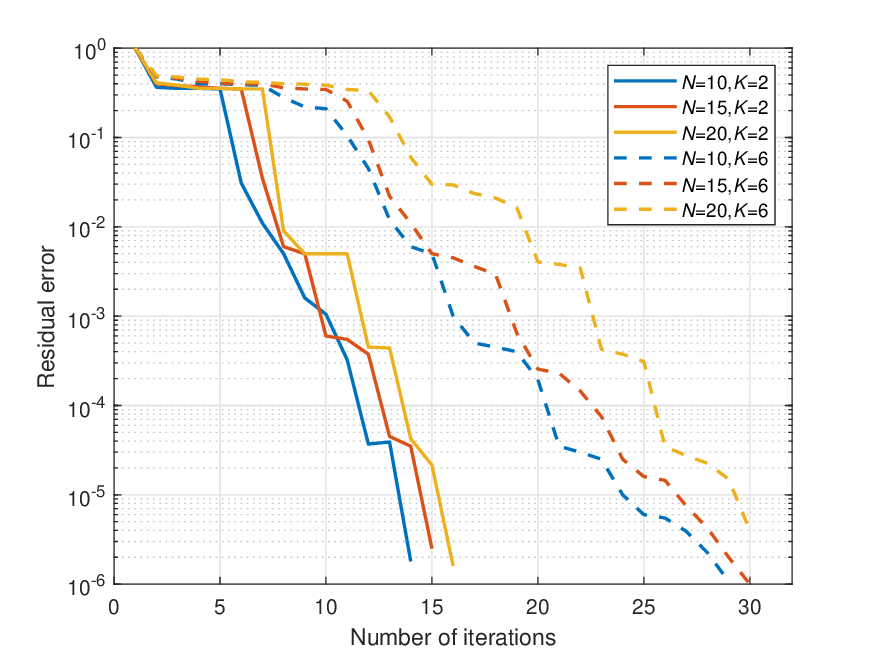}
	\end{center}
	\caption{Convergence rate of the proposed Algorithm 3, where ${\Gamma_{p}}=8$ dB, $\epsilon_3=10^{-6}$, and ${P_\text{Tol}}=10$ dBm.}
	\label{fig5}
\end{figure}

Figure \ref{fig5} presents the convergence rate of Algorithm 3. The barrier control parameters are set as $\varsigma_0=20$ and $\ell=2$, respectively. The initial values in Algorithm 3 are randomly generated. A general observation in Fig. \ref{fig5} is that Algorithm 3 presents a fast convergence rate. 
Besides, the number of iterations slightly increases as the number of transmit antennas and IoDs increases. It should be mentioned that for current hardware platforms, the 200MHz clock rate can be easily implemented \cite{hardware}. Therefore, it has enough time to calculate and update variables, and thus realize beam tracking for mobility of IoDs in practice.

We should mention that the confidential message can be correctly decoded only if both the recover of common parts and the private parts are correct. It is required that the received SINR of the private stream is over the prescribed minimum received value, or else system secrecy sum-rate (SSR) is zero. We define the average system SSR as
\begin{eqnarray}
	{C_{{\rm{Sys}}}} = \left\{ {\begin{array}{*{20}{c}}
			{{{\left[ {\sum\limits_{k \in  \mathcal{K}} {\left( {R_{c,k}^u - \mathop {\max }\limits_{q \in \mathcal{Q}} {\min \{ R_{c,q}^e,R_{p,q,k}^e\} }} \right)} } \right]}^ + }},\\\kern 120pt {\rm{if}} \kern 4pt{\gamma _{p,k}^u \ge {\Gamma _p}},\\
			\kern 56pt	0,\kern 56pt  {\rm{if}}\kern 4pt {\gamma _{p,k}^u < {\Gamma _p}}.
	\end{array}} \right.
	\label{eq42}\end{eqnarray}
\begin{figure}[tb]
	\begin{center}
		\includegraphics[width=1\columnwidth]{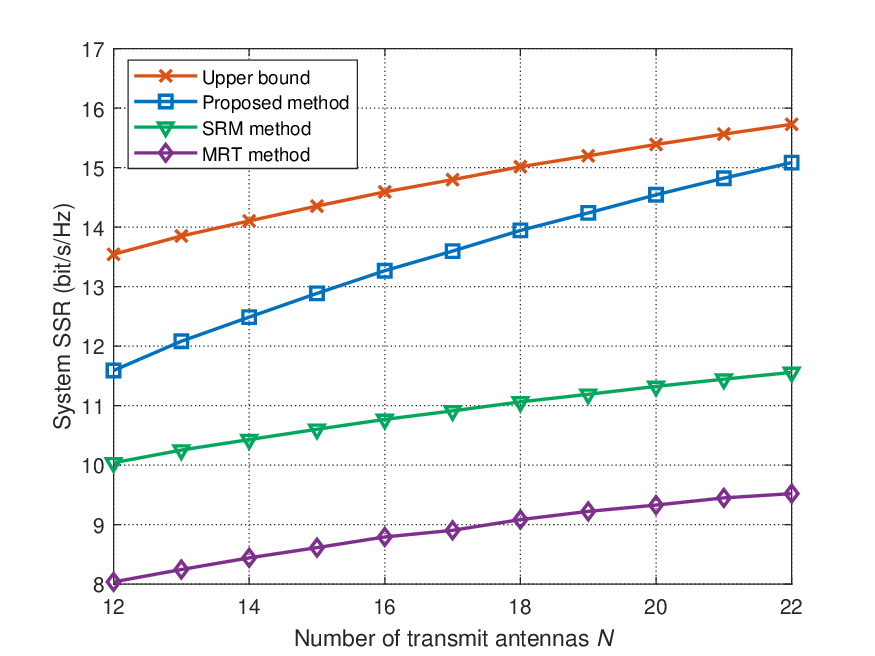}
	\end{center}
	\caption{The system SSR versus the number of transmit antennas for different method, where ${\Gamma_{p}}=8$ dB and ${P_\text{Tol}}=10$ dBm.}
	\label{fig6}
\end{figure}
\begin{figure}[tb]
	\begin{center}
		\includegraphics[width=1\columnwidth]{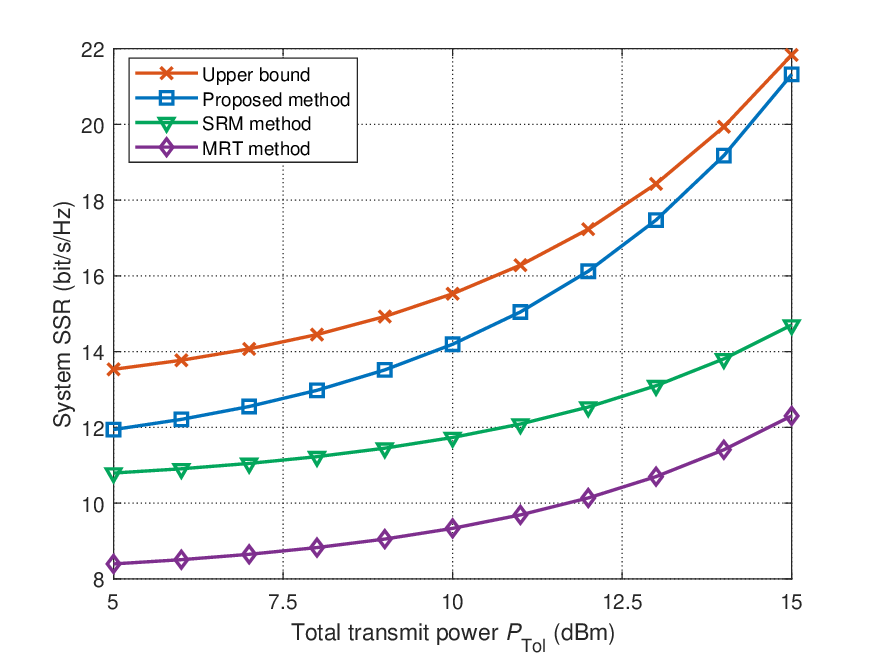}
	\end{center}
	\caption{The system SSR versus the total transmit power for different method, where ${\Gamma  _{p}}=8$ dB and $N=20$.}
	\label{fig7}
\end{figure}
\begin{figure}[tb]
	\begin{center}
		\includegraphics[width=1\columnwidth]{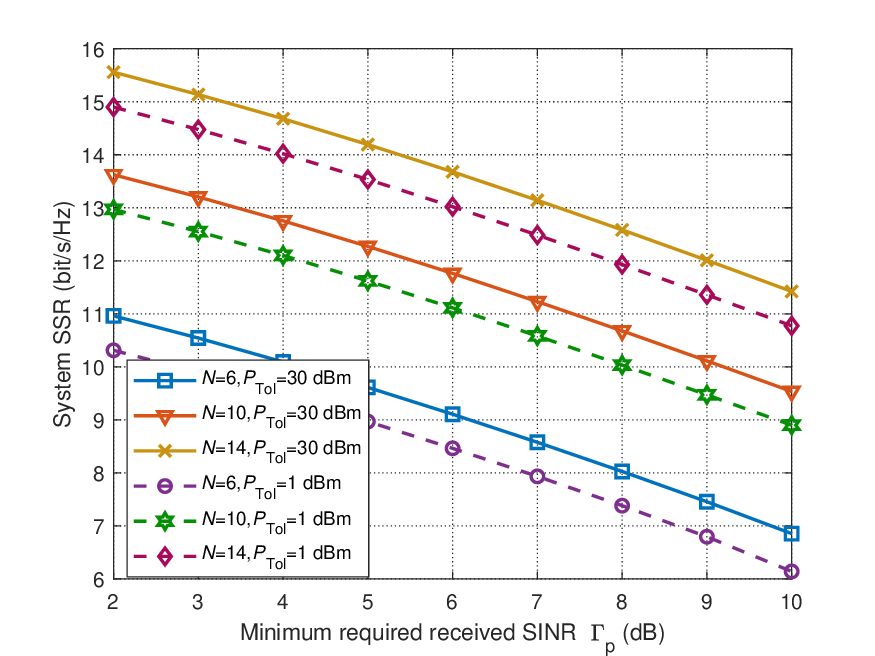}
	\end{center}
	\caption{The system SSR versus the minimum required received SINR of the private message for different system configurations.}
	\label{fig8}
\end{figure}
\begin{figure}[tb]
	\begin{center}
		\includegraphics[width=1\columnwidth]{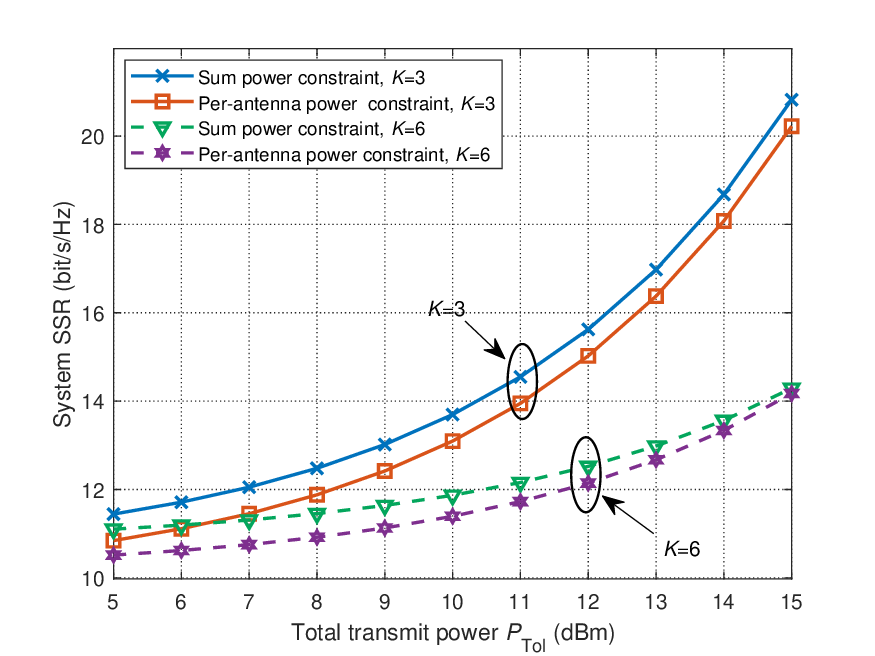}
	\end{center}
	\caption{The system SSR versus the total transmit power for different types of power constraints and number of IoD, where ${\Gamma  _{p}}=8$ dB and $N=20$.}
	\label{fig9}
\end{figure}

Then, we illustrate the average system SSR (bit/s/Hz) versus the number of antennas in Fig. \ref{fig6}. For a fair comparison, the total transmit power is set to equal. In addition, we also show the secrecy rate upper bound, i.e., the maximum IoD achievable rate under no passive Eve existence in the system. We can observe that the average system SSR increases as the number of transmit antennas increases. And by all accounts in array signal processing, more antennas equipped with the transmit array enhance the array’s capability in the degree of spatial freedom. Especially, the base station is more efficient beamforming for the information and AN with more transmit antennas. Moreover, the gap between the average system SSR and the secrecy rate upper bound diminishes when the number of transmit antennas is relatively large. Under our configurations, it is apparent that the secure performance of proposed scheme is better than other baseline schemes. The superior secrecy performance comes from the dual-use of the common messages and the enveloped private messages. 

In Fig. \ref{fig7}, we present the average system SSR versus the total transmit power. As expected, a better  performance is yielded for the proposed scheme compared with other baseline methods. More particularly, our design gradually converges to theoretic secrecy rate upper bound. By observing in Fig. \ref{fig7}, we can also see that the average system SSR shows an increasing trend with total transmit power. The performance gain obtained by higher IoD achievable rate and lower Eve rate is due to more transmit power utilization.

In Fig. \ref{fig8}, we explore the effect of system configurations for proposed method on average system SSR. This evaluation confirms that the average system SSR drops as the minimum required received SINR of the private stream increases. It is true since more transmit power is allocated to the private stream when the minimum required received SINR of the private stream becomes more stringent. Interestingly, a higher total transmit power results in a better average system SSR. Besides, it is also shown that the average system SSR increases with the increasing number of antennas.

In the last scenario, we quantify two different types of power constraints, i.e., sum power constraint and per-antenna power constrain. As shown in Fig. \ref{fig9}, the sum power constraint causes better secure performance compared to the per-antenna power constrain. The gaps between
the two types of power constraints are negligible in large number and high transmit power. This is because large number of IoDs and high transmit power are short of the DoF to take advantage of multiuser diversity.

\section{Conclusion}
In this paper, an AN-aided RS-based beamforming scheme was proposed to enhance the PHY security over MU-MISO IoT systems in the presence of passive Eves. We developed a F-SSR maximization problem by jointly optimizing transmit information and AN beamforming while satisfying per-antenna power constraints and prescribed minimum received SINR of the private stream. To solve this challenging problem, we studied a two-stage algorithm. More specifically, we first facilitated the non-concave parts simplification by the lower bound constraints with fixed allowable SINR, and then SDP relaxation method was adopted to solve the reformulated problem. Next, the BFGS algorithm was developed to search the global optimal solution. Additionally, an efficient algorithm was developed to solve the mimi-max program. Finally, simulation results demonstrated the superiority of the proposed scheme to provide PHY security in IoT communications. More importantly, the proposed scheme is not limited to only IoT networks. It is also appropriate in many applications requiring a high-level of security, such as satellite communications, unmanned aerial vehicles networks, military communications, millimeter-Wave communications.

\appendices
\section{Proof of Lemma 1}
The left-hand side in constraint \eqref{eq13b} can be equal to
\begin{eqnarray}
\Pr \left( {\mathop {\max }\limits_{q \in \mathcal{Q}} \gamma _{p,q,k}^e \le {\Gamma _e}} \right)\! =\! \prod\limits_{q \in \mathcal{Q}} \Pr \left( \gamma _{p,q,k}^e  \le {\Gamma _e} \right), \forall k \in \mathcal{K}.
\label{eq43}\end{eqnarray}
Then, by some mathematical manipulations, it is rearranged as
\begin{eqnarray}
	\begin{aligned}[b]
	&\Pr \left( {\mathop {\max }\limits_{q \in {\cal Q}} \gamma _{p,q,k}^e \le {\Gamma _e}} \right) \ge \kappa \\&
	\Leftrightarrow \Pr \left\{ {{\rm{Tr}}\left( {{{\bf{H}}_{e,q}}{\bf{A}}}\right) \le {\Gamma _e}\sigma _e^2} \right\} \ge {\kappa ^{1/Q}},
\end{aligned}
\label{eq44}\end{eqnarray}
where ${{\bf{H}}_{e,q}} \buildrel \Delta \over = {{\bf{h}}_{e,q}}{\bf{h}}_{e,q}^H$, and ${\bf{A}} \buildrel \Delta \over =  {{\bf{W}}_k} - {\Gamma _e}\sum\limits_{i \in {\mathcal{K}_c}\backslash k} {{{\bf{W}}_i}}  - {\Gamma _e}{{\bf{V}}_0}{\bf{BV}}_0^H $. Note that the Eves are modeled equivalent random channels following i.i.d., and thus the index of Eve channel can be removed.
 
Then, the probabilistic constraint upper bound can be expressed as
\begin{eqnarray}
	\begin{aligned}[b]
	{\mathop{\rm Tr}\nolimits} \left( {{{\bf{H}}_e}{\bf{A}}} \right) & \overset{{(a)}}\le \sum\limits_{i = 1}^N {{\lambda _i}} \left( {{{\bf{H}}_e}} \right){\lambda _i}({\bf{A}})\\&
    \overset{{(b)}}={\lambda _{\max }}\left( {{{\bf{H}}_e}} \right){\lambda _{\max }}({\bf{A}})\\&
	\overset{(c)}  =  {\mathop{\rm Tr}\nolimits} \left( {{{\bf{H}}_e}} \right){\lambda _{\max }}({\bf{A}}),
\end{aligned}
\label{eq45}\end{eqnarray}
where ${{\lambda _i}}\left( \cdot  \right)$, ${{\lambda _{\max }}}\left( \cdot  \right)$, and ${{\lambda _{\min }}}\left( \cdot  \right)$ indicate the $i$th, maximum, and minimum eigenvalue of the matrix, respectively, and the order satisfies ${\lambda _{\max }}\left(  \cdot  \right) = {\lambda _1}\left(  \cdot  \right) \ge {\lambda _i}\left(  \cdot  \right) \ge ... \ge {\lambda _N}\left(  \cdot  \right) = {\lambda _{\min }}\left(  \cdot  \right)$. Additionally, the inequality $(a)$ holds for the trace inequality of the positive Hermitian matrices \cite{Inequalities_Marshall}. The equations $(b)$ and $(c)$ can be established since  ${{{\bf{H}}_e}}$ is a rank-one positive semidefinite matrix.

According to \eqref{eq43}, \eqref{eq44}, and \eqref{eq45}, we get
\begin{align}
\Pr \left\{ {{\rm{Tr}}\left( {{{\bf{H}}_e}{\bf{A}}} \right) \le {\Gamma _e}\sigma _e^2} \right\} \ge \Pr \left\{ {{\rm{Tr}}\left( {{{\bf{H}}_e}} \right){\lambda _{\max }}\left( {\bf{A}} \right) \le {\Gamma _e}\sigma _e^2} \right\}.
\label{eq46}\end{align}

As a result, the probabilistic constraint can be converted as
\begin{eqnarray}
	\begin{aligned}[b]
	&\Pr \left\{ {\mathop {\max }\limits_{q \in {{\cal Q}}} \gamma _{p,q,k}^e \le {\Gamma _e}} \right\}\\&
	\ge \Pr \left\{ {{\rm{Tr}}\left( {{{\bf{H}}_e}} \right){\lambda _{\max }}\left( {\bf{A}} \right) \le {\Gamma _e}\sigma _e^2} \right\} \ge {\kappa ^{1/Q}}\\&
	\overset{{(d)}} \Leftrightarrow \Pr \left\{ {\frac{{{\lambda _{\max }}\left( {\bf{A}} \right)}}{{{\Gamma _e}\sigma _e^2}} \le \frac{1}{{{\rm{Tr}}\left( {{{\bf{H}}_e}} \right)}}} \right\} \ge {\kappa ^{1/Q}}\\&
	\overset{{(e)}} \Leftrightarrow \Pr \left\{ {\frac{1}{{{\rm{Tr}}\left( {{{\bf{H}}_e}} \right)}} \le \frac{{{\lambda _{\max }}\left( {\bf{A}} \right)}}{{{\Gamma _e}\sigma _e^2}}} \right\} \le 1 - {\kappa ^{1/Q}}\\&
	\overset{{(f)}} \Leftrightarrow {\lambda _{\max }}\left( {\bf{A}} \right) \le \Phi _N^{ - 1}\left( {1 - {\kappa ^{1/Q}}} \right){\Gamma _e}\sigma _e^2\\&
	\Leftrightarrow {\bf{A}} \preceq {{\bf{I}}_N}\left[ {\Phi _N^{ - 1}\left( {1 - {\kappa ^{1/Q}}} \right){\Gamma _e}\sigma _e^2} \right].
\end{aligned}
\label{eq47}\end{eqnarray}
The equivalent transformation $(d)$ can be got due to positive definite matrix ${{\bf{H}}_e}$, $(e)$ holds for a basic property of probability, and $(f)$ follows steps similar to \cite[Lemma 2]{WIPT_Ng}. Note that the implication can also be applied to any continuous channel distribution by replacing  $\Phi _N^{ - 1}\left(  \cdot  \right)$ with an inverse c.d.f. with respect to the corresponding distribution. Thus, Lemma 1 follows.

\section{Proof of Proposition 2}
Utilizing the following lemma,

\textsl{Lemma A} \cite[Coroll. 2]{convergence_Grippo}: Given the problem
\begin{eqnarray}
\mathop {\min }\limits_{{\bf{W}},\chi}  {\kern 4pt} \mathscr{F} \left({\bf{W}},\chi \right){\mkern 1mu} {\kern 20pt} {\rm{s.t.}}{\mkern 1mu} {\kern 2pt} \left({\bf{W}},\chi \right) \in {\cal W} \times {\cal A},
\label{eq48}\end{eqnarray}
where $ \mathscr{F} \left({\bf{W}},\chi \right)$ is a continuously differentiable function, and ${\cal W} \subseteq \mathbb{C}$ and ${\cal A} \subseteq \mathbb{R}$ are closed, nonempty, and convex subsets, every limit point of the iterates is a stationary point.

Then, problem \eqref{eq21} is rewritten as 
\begin{align}
	&\mathop {\max }\limits_{{{\bf{W}}_i},{\bf{B}}}  \kern 2pt\sum\limits_{k \in \mathcal{K}} {{{\log }_2}} \left( {{\bf{h}}_{u,k}^H{\bf{W}}_c{{\bf{h}}_{u,k}}\! +\! \sum\limits_{i \in \mathcal{K}} {{\bf{h}}_i^H{\bf{W}}_i{{\bf{h}}_i}} \! +\! \sigma _u^2} \right)\notag \\
	&\kern 20pt \!+\!\sum\limits_{k \in {\cal K}}{\mathscr{G}_{\zeta_k }}\left(\chi_k \right)\notag \\
	&{\rm{s.t.}}  \kern 2pt  \eqref{eq15b},\eqref{eq13c},\eqref{eq13d}.
\label{eq49}\end{align}
The objective of \eqref{eq49} is continuously differentiable. The feasible set is closed, nonempty, and convex. By Bolzano–Weierstrass theorem, the sequence ${{\bf{Z}}^{(r)} }=( {\{ {{\bf{W}}_k^{(r)}} \}_{k \in \mathcal{K}},{\chi_k^{(r)}}} )$ updated by solving problem  \eqref{eq49} has limit points. Invoking Lemma A, we know that every limit point ${{{\bf{Z}}^ \star }}$ generated by Algorithm 1 is a stationary point of \eqref{eq49}. Then, we will prove that every stationary point of problem \eqref{eq49} is also a stationary point of \textbf{P4}. We use ${{ \mathscr{F}_1}\left( {{{\boldsymbol{{\rm Z}}}}} \right)}$ and ${ \mathscr{F}_2}\left( {{{\left\{ {{\bf{W}}_k} \right\}}_{k \in \mathcal{K}}}} \right)$ to denote the objective of problem \eqref{eq49} and \textbf{P4}, respectively.  ${{\bf{Z}}^ \star }$ is a stationary point of\eqref{eq49}, yielding
\begin{align}
&	{\mathop{\rm Tr}\nolimits} \left[ {{\nabla_{{{\bf{W}}_k}}}{ \mathscr{F} _1^H}{{\left( {{{\bf{{\boldsymbol Z}}}^ \star }} \right)} }\left( {{{\bf{W}}_k}\! -\! {\bf{W}}_k^ \star } \right)} \right]\! \le\! 0,{\forall k \in \mathcal{K}}, \label{eq50}\\
&	{\nabla _\chi }{ \mathscr{F} _1^H}{\left( {{{\bf{Z}}^ \star }} \right) }\left( {\chi_k  - {\chi_k ^ \star }} \right) \le 0,\forall \chi_k  > 0.
	\label{eq51}\end{align}
According to Proposition 1, \eqref{eq18},  and \eqref{eq51}, we obtain
\begin{eqnarray}
	{\chi_k ^{\star}} = {\left( {\sum\limits_{i \in \mathcal{K}} {{\bf{h}}_{u,k}^H{\bf{W}}_i^\star} {{\bf{h}}_{u,k}} + \sigma _u^2} \right)^{ - 1}}.
	\label{eq52}\end{eqnarray}
Inserting \eqref{eq52} into \eqref{eq50}, one can easily verify that
\begin{eqnarray}
	{\nabla _{{{\bf{W}}_k}}}{\mathscr{F}_2}\left( {{{\left\{ {{\bf{W}}_k^ \star } \right\}}_{k \in\mathcal{K}}}} \right) = {\nabla_{{{\bf{W}}_k}}}{\mathscr{F}_1}\left( {{{\bf{Z}}^ \star }} \right).
\label{eq53}\end{eqnarray}
Based on  \eqref{eq50} and \eqref{eq53}, we thus claim that
\begin{align}
	&{\mathop{\rm Tr}\nolimits} \left[ {{\nabla _{{{\bf{W}}_k}}}{ \mathscr{F} _2^H}\left( {{{\left\{ {{\bf{W}}_k^ \star } \right\}}_{k \in \mathcal{K}}}} \right)\left( {{{\bf{W}}_k} - {\bf{W}}_k^ \star } \right)} \right] \le 0,{\forall k \in \mathcal{K}} \notag\\
	& {\rm{s.t.}}  \kern 2pt \eqref{eq15b},\eqref{eq13c},\eqref{eq13d}.
\label{eq54}\end{align}
In other words, ${{{\left\{ {{\bf{W}}_k^ \star } \right\}}_{k \in \mathcal{K}}}}$ is the optimal solution of the following problem
\begin{align}
	\mathop {{\rm{max}}}\limits_{{{\left\{ {{{\bf{W}}_k}} \right\}}_{k \in  \mathcal{K}}}} \kern 2pt{\rm{Tr}}\left[ {{\mathscr F}_2^H\left( {{{\left\{ {{\bf{W}}_k^ \star } \right\}}_{k \in  \mathcal{K}}}} \right)} \right] \kern 4pt {\rm{s.t.}}  \kern 2pt  \eqref{eq15b},\eqref{eq13c},\eqref{eq13d}.
	\label{eq55}\end{align}
Consequently, the conditions in \eqref{eq55} are exactly the KKT conditions in \textbf{P4}, and thus, proposition 2 follows.

\section{Proof of Theorem 1}
The dual function is the minimum value of the Lagrangian function \textbf{P6} over ${{\boldsymbol{\Psi }}_k}$, i.e., \cite [CH.5] {Convex_Boyd}
\begin{align}
	&{\mathcal D}_1({{\boldsymbol{\Psi }}_k})\notag\\&
	= \mathop {\inf }\limits_{{\Gamma _e} > 0} {{{\cal L}}_1}\left( {{\Gamma _e},{{\bf{\Psi }}_k}} \right)\notag\\&
	=\mathop {\inf }\limits_{{\Gamma _e} > 0} \left\{ {{{\log }_2}\left( {{\Gamma _e}} \right) \!+\! \sum\limits_{k \in {\cal K}} {{\rm{Tr}}\left( {{{\bf{\Psi }}_k}{{\bf{\Pi }}_k}} \right)} {\Gamma _e}} \right\} \! - \!\sum\limits_{k \in {\cal K}} {{\rm{Tr}}\left( {{{\bf{\Psi }}_k}{{\bf{W}}_k}} \right)} \notag\\&
	= \!  -\! \mathop {\sup }\limits_{{\Gamma _e} > 0} \left\{ {\! - \!{{\log }_2}\left( {{\Gamma _e}} \right) \!-\! \sum\limits_{k \in {\cal K}} {{\rm{Tr}}\left( {{{\bf{\Psi }}_k}{{\bf{\Pi }}_k}} \right)} {\Gamma _e}} \right\}\! -\! \sum\limits_{k \in {\cal K}} {{\rm{Tr}}\left( {{{\bf{\Psi }}_k}{{\bf{W}}_k}} \right)}\notag\\&
	= \! -\! {\mathscr{{ C}}^*}\left( { \!-\! \sum\limits_{k \in {\cal K}} {{\rm{Tr}}\left( {{{\bf{\Psi }}_k}{{\bf{\Pi }}_k}} \right)} } \right)\!  -\! \sum\limits_{k \in {\cal {\cal K}}} {{\rm{Tr}}\left( {{{\bf{\Psi }}_k}{{\bf{W}}_k}} \right)},
	\label{eq56}\end{align}
where ${\mathscr{C}}^*\left(  \cdot  \right)$ is the conjugate of ${\mathscr{C}}={\rm{log}}_2\left(  \cdot  \right)$, satisfying $ - {{\rm{log}}_2}\left( \gamma  \right)\overset{{{\mathscr{C}}^*\left(  \cdot  \right)}} \to  - {{\rm{log}}_2}\left( { - \gamma } \right) - 1$ \cite{Conjugate_Fenchel}. Therefore, the associated dual problem is given by
\begin{align}
	{\mathcal D}_1({{\boldsymbol{\Psi }}_k}) =\! - \!{\log _2}\left( { \!-\! \sum\limits_{k \in {\cal K}} {{\rm{Tr}}\left( {{{\bf{\Psi }}_k}{{\bf{\Pi }}_k}} \right)} } \right) \! -\! \sum\limits_{k \in {\cal K}}  {{\rm{Tr}}\left( {{{\bf{\Psi }}_k}{{\bf{W}}_k}} \right)}\! -\! 1.
	\label{eq57}\end{align}
Thus, Theorem 1 follows.

\section{Proof of Theorem 2}
The partial Lagrangian function of the problem \textbf{P7} is expanded as
\begin{align}
	& \mathcal{L}_2\left( {{{\bf{W}}_c},{{\left\{ {{{\bf{W}}_k}} \right\}}},{\bf{B}},{{\left\{ {\bf{D}}_k \right\}}},{{\left\{ {\lambda _n} \right\}}},{{\left\{ {\mu _k} \right\}}},{{\left\{ {\upsilon_k} \right\}}} } \right)\notag\\&
	= \sum\limits_{k \in {\cal K}} {{{\log }_2}\left( {1\! +\! {\bf{h}}_{u,k}^H{{\bf{W}}_c}{{\bf{h}}_{u,k}}} \right)}\notag \\&
	\kern 0pt \!-\! \sum\limits_{k \in {\cal K}} {{\rm{Tr}}\left\{ {{{\bf{D}}_k}\left( {{{\bf{W}}_k}\! -\! {\Gamma _e}{\sum _{i \in {{\cal K}_c}\backslash k}}{{\bf{W}}_i}\! -\! {\Gamma _e}{{\bf{V}}_0}{\bf{BV}}_0^H\! -\! {{\bf{I}}_N}\xi } \right)} \right\}}\notag \\&
	\kern 0pt \!-\! \sum\limits_{n \in {\cal N}} {{\lambda _n}\left[ {\sum\limits_{i \in {{\cal K}_c}} {{\rm{Tr}}\left( {{{\bf{W}}_i}{{\bf{E}}^{(n)}}} \right)} \! + \!{\rm{Tr}}\left( {{\bf{B}}{{{\bf{\bar E}}}^{(n)}}} \right) \!-\! {P_n}} \right]}\notag \\&
	\kern 0pt \!-\! \sum\limits_{k \in {\cal K}} {{\mu _k}\left[ {{\Gamma _{p}}\sum\limits_{i \in {{\cal K}}\backslash k} {{\rm{Tr}}\left( {{{\bf{H}}_{u,k}}{{\bf{W}}_i}} \right)} \! +\! {\Gamma _{p}}\sigma _{u}^2 \!-\! {\rm{Tr}}\left( {{{\bf{H}}_{u,k}}{{\bf{W}}_k}} \right)} \right]}\notag \\&
	\kern 0pt \!-\! \sum\limits_{k \in {\cal K}} {{\upsilon _k}\left( {\sum\limits_{i \in {\cal K}} {{\rm{Tr}}\left( {{{\bf{H}}_{u,k}}{{\bf{W}}_i}} \right)} \! +\! \sigma _u^2 \!-\! {\Sigma _p}} \right)},
	\label{eq58}\end{align}
where ${{\bf{H}}_{u,k}} \buildrel \Delta \over = {{\bf{h}}_{u,k}}{\bf{h}}_{u,k}^H$.

We use a given set $\left( {{\bf{D}}_k,{\boldsymbol{\lambda }},{\boldsymbol{\mu }},{\boldsymbol{\upsilon }}} \right)$ to transform the partial Lagrangian function \eqref{eq58} into
\begin{align}
	&\mathcal{L}_2\left( {{{\bf{W}}_c},{{\left\{ {{{\bf{W}}_k}} \right\}}},{\bf{B}},{{\left\{ {\bf{D}}_k \right\}}},{{\left\{ {\lambda _n} \right\}}},{{\left\{ {\mu _k} \right\}}},{{\left\{ {\upsilon_k} \right\}}} } \right)\notag\\&
	= \sum\limits_{k \in {\cal K}} {{{\log }_2}\left( {1 + {\bf{h}}_{u,k}^H{{\bf{W}}_c}{{\bf{h}}_{u,k}}} \right)}  - {\rm{Tr}}\left( {{\bf{\Sigma }}{{\bf{W}}_c}} \right)\notag \\&
	\kern 10pt- \sum\limits_{k \in {\cal K}} {{\rm{Tr}}\left( {{{\bf{\Omega }}_k}{{\bf{W}}_k}} \right)}  - {\rm{Tr}}\left( {{\bf{\Phi B}}} \right) + \xi \sum\limits_{k \in {\cal K}} {{\rm{Tr}}\left( {{{\bf{D}}_k}} \right)} \notag\\&
	\kern 10pt	+ {{\bf{p}}^T}{\boldsymbol{\lambda }} - {\bf{d}}^T{\boldsymbol{\mu }} - {\bf{b}}^T{\boldsymbol{\upsilon }},
	\label{eq59}\end{align}
where ${\bf{\Sigma }} = {\rm{diag}}({\boldsymbol{\lambda }}) - {\Gamma _e}\sum\limits_{k \in {\cal K}} {{{\bf{D}}_k}}$, ${{\bf{\Omega }}_k} = {{\bf{D}}_k} + {\rm{diag}}({\boldsymbol{\lambda }}) - {\mu _k}{{\bf{H}}_{u,k}} + {\upsilon _k}{{\bf{H}}_{u,k}} + \sum\limits_{i \in {\cal K}\backslash k} {\left( {{\mu _i}{\Gamma _{p,i}}{{\bf{H}}_{u,i}} + {\upsilon _i}{{\bf{H}}_{u,i}} - {\Gamma _e}{{\bf{D}}_i}} \right)}$, ${\bf{\Phi }} = \sum\limits_{n \in \cal N} {\lambda _n}{{{\bf{\bar E}}}^{(n)}} - \sum\limits_{k \in {\cal K}} {\Gamma _e}{{\bf{V}}_0}{{\bf{D}}_k}{\bf{V}}_0^H$, ${\boldsymbol{\lambda }} = [ {\lambda _1},{\lambda _2},...,{\lambda _N} ]^T$, ${\boldsymbol{\mu }} = [ {\mu _1},{\mu _2},...,{\mu _K} ]^T$, ${\boldsymbol{\upsilon }} = [ {\upsilon _1},{\upsilon _2},...,{\upsilon _K} ]^T$, ${\bf{p}}=[ {P_1}$, ${P_2}$, $ ...$, ${P_N} ]^T$, ${\bf{d}}={{\Gamma _p}\sigma _u^2}{\boldsymbol{1}}^T$, and ${\bf{b}} = ( {\sigma _u^2 - {\Sigma _p}} ){\boldsymbol{1}}^T$.

For simplicity, we define ${{{\bf{\bar W}}}_c} \buildrel \Delta \over = {{\bf{\Sigma }}^{1/2}}{{\bf{W}}_c}{{\bf{\Sigma }}^{1/2}}$. Thereby, the dual objective of \textbf{P7} is established by
\begin{align}
	& \mathcal{D}_2\left( {\bf{D}}_k,{\boldsymbol{\lambda }},{\boldsymbol{\mu }},{\boldsymbol{\upsilon }} \right)\notag \\
	&= \mathop {\max }\limits_{{{\bf{\bar W}}_c}, {{\bf{W}}_k},{\bf{B}}} \kern 2pt {\mathcal{L}}_2 \left( {{{\bf{\bar W}}_c},{{\left\{ {{{\bf{W}}_k}} \right\}}},{\bf{B}},\left\{{\bf{D}}_k\right\},{\boldsymbol{\lambda }},{\boldsymbol{\mu }},{\boldsymbol{\upsilon }}} \right).
	\label{eq60}\end{align}

We use the results developed in \cite[Appendix A]{Duality_Vishwanath} to transform \eqref{eq60} into
\begin{align}
	&{\mathcal D}_2\left( {{{\bf{D}}_k},{\boldsymbol{\lambda }},{\boldsymbol{\mu }},{\boldsymbol{\upsilon }}} \right)\notag\\&
	= \mathop {\max }\limits_{\omega  \ge 0,{\bf{B}} \succeq \boldsymbol{0}} \sum\limits_{k \in {\cal K}} {{{\log }_2}\left| {{\bf I} + {{\bf{\Sigma }}^{ - 1/2}}{{\bf{h}}_{u,k}}\omega {\bf{h}}_{u,k}^H{{\bf{\Sigma }}^{ - 1/2}}} \right|}  - \omega \notag\\&
	\kern 10pt - \sum\limits_{k \in {\cal K}} {{\rm{Tr}}\left( {{{\bf{\Omega }}_k}{{\bf{W}}_k}} \right)}  - {\rm{Tr}}\left( {{\bf{\Phi B}}} \right) + \xi \sum\limits_{k \in {\cal K}} {{\rm{Tr}}\left( {{{\bf{D}}_k}} \right)}\notag\\&
	\kern 10pt + {{\bf{p}}^T}{\boldsymbol{\lambda }} -  {\bf{d}}^T{\boldsymbol{\mu }} -  {\bf{b}}^T{\boldsymbol{\upsilon }},
	\label{eq61}\end{align}
where the relation between $\omega$ in \eqref{eq61} and ${{{\bf{\bar W}}}_c}$ in \eqref{eq60} satisfies
\begin{align}
&\omega  = \frac{1}{{\left\| {{{\bf{\Sigma }}^{ - 1/2}}{{\bf{h}}_{u,k}}} \right\|_2^2}}{\bf{h}}_{u,k}^H{{\bf{\Sigma }}^{ - 1/2}}{{{\bf{\bar W}}}_c}{{\bf{\Sigma }}^{ - 1/2}}{{\bf{h}}_{u,k}},	\label{eq62}\\&
{{{\bf{\bar W}}}_c} = \frac{1}{{\left\| {{{\bf{\Sigma }}^{ - 1/2}}{{\bf{h}}_{u,k}}} \right\|_2^2}}{{\bf{\Sigma }}^{ - 1/2}}{{\bf{h}}_{u,k}}\omega {\bf{h}}_{u,k}^H{{\bf{\Sigma }}^{ - 1/2}}.
	\label{eq63}\end{align}
Then, \eqref{eq61} can be recast into a more compact form by
\begin{align}
	&{\cal D}_2\left( {{{\bf{D}}_k},{\bf{g}}} \right)\notag\\&
	= \mathop {\max }\limits_{\omega  \ge 0,{\bf{B}} \succeq \boldsymbol{0}}\kern 2pt \sum\limits_{k \in {\cal K}} {{{\log }_2} {\frac{\left|{{\bf{\Sigma }}\! +\! {{\bf{h}}_{u,k}}\omega {\bf{h}}_{u,k}^H} \right|}{\left|{\bf{\Sigma }}\right|}}} \! -\! \omega\! -\! \sum\limits_{k \in {\cal K}} {{\rm{Tr}}\left( {{{\bf{\Omega }}_k}{{\bf{W}}_k}} \right)}\notag \\&
	\kern 10pt - {\rm{Tr}}\left( {{\bf{\Phi B}}} \right) + \xi \sum\limits_{k \in {\cal K}} {{\rm{Tr}}\left( {{{\bf{D}}_k}} \right)}  + {{\bf{g}}^T}{\boldsymbol{\eta }},
	\label{eq64}\end{align}
where ${\bf{g}} = {\left[ {{{\bf{p}}^T},-{{\bf{d}}^T},-{{\bf{b}}^T}} \right]^T}$, and ${\boldsymbol{\eta }} = {\left[ {{{\boldsymbol{\lambda }}^T},{{\boldsymbol{\mu }}^T},{{\boldsymbol{\upsilon }}^T}} \right]^T}$.
The Lagrange dual problem can be simplified via minimizing ${\cal D}_2\left( {{{\bf{D}}_k},{\bf{g}}} \right)$, i.e.,
\begin{eqnarray}
\mathop {\min }\limits_{{{\bf{D}}_k} \succeq \boldsymbol{0},{\boldsymbol{\eta }} \ge \boldsymbol{0}} {\cal D}_2\left( {{{\bf{D}}_k},{\boldsymbol{\eta }}} \right)
\label{eq65}\end{eqnarray}
or an explicit subjective as
\begin{align}
&\mathop {\min }\limits_{{{\bf{D}}_k} \succeq \boldsymbol{0},{\boldsymbol{\eta }} \ge \boldsymbol{0}}\kern 4pt \mathop {\max }\limits_{\omega  \ge 0,{{\bf{W}}_k} \succeq \boldsymbol{0},{\bf{B}} \succeq \boldsymbol{0}} \sum\limits_{k \in {\cal K}} {{{\log }_2} {\frac{\left|{{\bf{\Sigma }}\! +\! {{\bf{h}}_{u,k}}\omega {\bf{h}}_{u,k}^H} \right|}{\left|{\bf{\Sigma }}\right|}}} - \omega \notag \\&
\kern 10pt- \sum\limits_{k \in {\cal K}} {{\rm{Tr}}\left( {{{\bf{\Omega }}_k}{{\bf{W}}_k}} \right)}  - {\rm{Tr}}\left( {{\bf{\Phi B}}} \right) + \xi \sum\limits_{k \in {\cal K}} {{\rm{Tr}}\left( {{{\bf{D}}_k}} \right)}  + {{\boldsymbol{g}}^T}{\boldsymbol{\eta }}.
	\label{eq66}\end{align}
Introducing an intermediate variable $z>0$, problem \eqref{eq66} is equivalently transformed as
\begin{align}
	&\mathop {\min }\limits_{{{\bf{D}}_k} \succeq \boldsymbol{0},{\boldsymbol{\eta }} \ge \boldsymbol{0}} \kern 4pt \mathop {\max }\limits_{\omega  \ge 0,{{\bf{W}}_k} \succeq \boldsymbol{0},{\bf{B}} \succeq \boldsymbol{0}} \sum\limits_{k \in {\cal K}} {{{\log }_2}\frac{{\left| {{\bf{\Sigma }} + {{\bf{h}}_{u,k}}\omega {\bf{h}}_{u,k}^H} \right|}}{{\left| {\bf{\Sigma }} \right|}}}  - z\vartheta  \notag\\&
	\kern 20pt + \xi \sum\limits_{k \in {\cal K}} {{\rm{Tr}}\left( {{{\bf{D}}_k}} \right)}  + {{\bf{g}}^T}{\boldsymbol{\eta }} \notag\\&
	{\rm{s.t.}} \kern 4pt	\omega +\sum\limits_{k \in {\cal K}} {{\rm{Tr}}\left( {{{\bf{\Omega }}_k}{{\bf{W}}_k}} \right)}  + {\rm{Tr}}\left( {{\bf{\Phi B}}} \right) \le z\vartheta. 
	\label{eq67}\end{align}
Then, we scale down to alter variables as
\begin{eqnarray}
\left\{ \begin{array}{l}
	\tilde \omega  = \omega /z,\\
	{\boldsymbol{\tilde \eta }} = {\boldsymbol{\eta }}/z,\\
	{{{\bf{\tilde D}}}_k} = {{\bf{D}}_k}/z,\\
	{\bf{\tilde B}} = {\bf{B}}/z,\\
	{{{\bf{\tilde W}}}_k} = {{\bf{W}}_k}/z,\\
	{\boldsymbol{\tilde \Sigma }} = {\boldsymbol{\Sigma }}/z.
\end{array} \right.
\label{eq68}\end{eqnarray}
Now, new optimization problem with respect $\tilde \omega$, ${\boldsymbol{\tilde \eta }}$, ${{{\bf{\tilde D}}}_k}$, $\bf{\tilde B}$ and	${{{\bf{\tilde W}}}}_k$ is given by
\begin{align}
	&\mathop {\min }\limits_{{{\bf{\tilde D}}_k} \succeq \boldsymbol{0},{\boldsymbol{\tilde \eta }} \ge 0}\kern 4pt \mathop {\max }\limits_{\tilde \omega  \ge 0,{{\bf{\tilde W}}_k} \succeq \boldsymbol{0},{\bf{\tilde B}} \succeq \boldsymbol{0}} \sum\limits_{k \in {\cal K}} {{{\log }_2}\frac{{\left| {{\bf{\tilde \Sigma }} + {{\bf{h}}_{u,k}}\tilde \omega {\bf{h}}_{u,k}^H} \right|}}{{\left| {{\bf{\tilde \Sigma }}} \right|}}} \notag\\&
	 \kern 20pt + z\left[ {\xi \sum\limits_{k \in {\cal K}} {{\rm{Tr}}\left( {{{{\bf{\tilde D}}}_k}} \right)}  + {{\bf{g}}^T}{\boldsymbol{\tilde \eta }} - \vartheta } \right]\notag\\&
	{\rm{s.t.}} \kern 4pt	\tilde \omega  + \sum\limits_{k \in {\cal K}} {{\rm{Tr}}\left( {{{\bf{\Omega }}_k}{{{\bf{\tilde W}}}_k}} \right)}  + {\rm{Tr}}\left( {{\bf{\Phi \tilde B}}} \right) \le \vartheta .
	\label{eq69}\end{align}
Obviously, we can form an equivalent problem by making these inequality constraints explicit, i.e., 
\begin{align}
	&\mathop {\min }\limits_{{{\bf{\tilde D}}_k} \succeq  \boldsymbol{0},{\boldsymbol{\tilde \eta }} \ge 0}\kern 4pt \mathop {\max }\limits_{\tilde \omega  \ge 0,{{\bf{\tilde W}}_k} \succeq  \boldsymbol{0},{\bf{\tilde B}} \succeq  \boldsymbol{0}} \sum\limits_{k \in  {\cal {\cal K}}} {{{\log }_2}\frac{{\left| {{\bf{\tilde \Sigma }} + {{\bf{h}}_{u,k}}\tilde \omega {\bf{h}}_{u,k}^H} \right|}}{{\left| {{\bf{\tilde \Sigma }}} \right|}}}  \notag\\&
	{\rm{s.t.}}\kern 4pt\tilde \omega  + \sum\limits_{k \in {\cal K}} {{\rm{Tr}}\left( {{{\bf{\Omega }}_k}{{{\bf{\tilde W}}}_k}} \right)}  + {\rm{Tr}}\left( {{\bf{\Phi \tilde B}}} \right) \le \vartheta, \notag\\&
	\kern 17pt \xi \sum\limits_{k \in {\cal K}} {{\rm{Tr}}\left( {{{{\bf{\tilde D}}}_k}} \right)}  + {{\bf{g}}^T}{\boldsymbol{\tilde \eta }} \le \vartheta,
	\label{eq70}\end{align}
Thus, Theorem 2 follows.

\ifCLASSOPTIONcaptionsoff
  \newpage
\fi

\bibliographystyle{IEEEtran}
\bibliography{IEEEabrv,REF}

\end{document}